\documentclass[aip,reprint]{revtex4-1}
\usepackage{graphicx}
\usepackage{dcolumn}
\usepackage{bm}

\usepackage[utf8]{inputenc}
\usepackage[T1]{fontenc}
\usepackage{mathptmx}
\usepackage{etoolbox}

\begin{document}


\title{An Integrated Quantum Material Testbed with Multi-Resolution Photoemission Spectroscopy} 



\author{Chenhui Yan}
\affiliation{Pritzker School of Molecular Engineering, University of Chicago, Chicago, Illinois 60637, USA}
\author{Emanuel Green}
\affiliation{Pritzker School of Molecular Engineering, University of Chicago, Chicago, Illinois 60637, USA}
\author{Riku Fukumori}
\affiliation{Pritzker School of Molecular Engineering, University of Chicago, Chicago, Illinois 60637, USA}
\author{Nikola Protic}
\affiliation{Pritzker School of Molecular Engineering, University of Chicago, Chicago, Illinois 60637, USA}
\author{Seng Huat Lee}
\affiliation{Department of Physics, Pennsylvania State University, University Park, State College, Pennslyvania 16802, USA}
\author{Sebastian Fernandez-Mulligan}
\affiliation{Pritzker School of Molecular Engineering, University of Chicago, Chicago, Illinois 60637, USA}
\author{Rahim Raja}
\affiliation{Pritzker School of Molecular Engineering, University of Chicago, Chicago, Illinois 60637, USA}
\author{Robin Erdakos}
\affiliation{Pritzker School of Molecular Engineering, University of Chicago, Chicago, Illinois 60637, USA}
\author{Zhiqiang Mao}
\affiliation{Department of Physics, Pennsylvania State University, University Park, State College, Pennslyvania 16802, USA}
\author{Shuolong Yang}
\email{yangsl@uchicago.edu}
\affiliation{Pritzker School of Molecular Engineering, University of Chicago, Chicago, Illinois 60637, USA}

\date{\today}

\begin{abstract}
We present the development of a multi-resolution photoemission spectroscopy (MRPES) setup which probes quantum materials in energy, momentum, space, and time. This versatile setup integrates three light sources in one photoemission setup, and can conveniently switch between traditional angle-resolved photoemission spectroscopy (ARPES), time-resolved ARPES (trARPES), and micron-scale spatially resolved ARPES ($\mu$ARPES). It provides a first-time all-in-one solution to achieve an energy resolution $< 4$~meV, a time resolution $< 35$~fs, and a spatial resolution $\sim 10$~$\mu$m in photoemission spectroscopy. Remarkably, we obtain the shortest time resolution among the trARPES setups using solid-state nonlinear crystals for frequency upconversion. Furthermore, this MRPES setup is integrated with a shadow-mask assisted molecular beam epitaxy system, which transforms the traditional photoemission spectroscopy into a quantum device characterization instrument. We demonstrate the functionalities of this novel quantum material testbed using FeSe/SrTiO$_3$ thin films and MnBi$_4$Te$_7$ magnetic topological insulators.
\end{abstract}

\maketitle 

\section{Introduction}
Angle-resolved photoemission spectroscopy (ARPES) has been established as a powerful tool to directly reveal the single-particle spectral function $A(\mathbf{k},\omega)$, which encodes electronic band structures and many-body interactions~\cite{Damascelli2003, Sobota2021}. ARPES has played a major role in the discovery of the $d$-wave superconducting gap in cuprate superconductors~\cite{Shen1993, Vishik2012}, the Dirac surface state in topological insulators~\cite{Hsieh2008,Chen2009}, and recently the putative topological superconducting gap in FeTe$_{0.55}$Se$_{0.45}$.~\cite{zhang2018}

Investigations on quantum materials often require the combination of different modalities in photoemission spectroscopy. Current solutions for multi-modal ARPES rely on the utilization of facilities at different geographical locations. For instance, one may use a laboratory-based ARPES setup to measure superconducting gaps in monolayer FeSe/SrTiO$_3$,~\cite{Liu2012, tan2013, he2013} but will need micro ($\mu$)-~\cite{Iwasawa2017} or nanoARPES~\cite{Barbo2000, Avila2013, Rotenberg2014} at specialized synchrotron end stations to understand the microscopic spatial variations~\cite{Faeth2021}. One may use a time-resolved ARPES (trARPES) setup~\cite{Perfetti2007, Schmitt2008, Sobota2012, Smallwood2012, Wang2013, Yang2015inequivalence, Gerber2017}  to resolve the dynamics of unoccupied electronic states in a topological insulator Bi$_{2}$Se$_{3}$~\cite{Sobota2012, Sobota2013}, but will need high-photon-energy synchrotron-based measurements to reveal the valence band dispersions~\cite{Soifer2019}. Importantly, multi-modal ARPES measurements are remarkably challenging due to the limited access to multiple facilities and the potential sample degradations during transportation.

\begin{figure}
    \centering
    \includegraphics[width = 3.5 in]{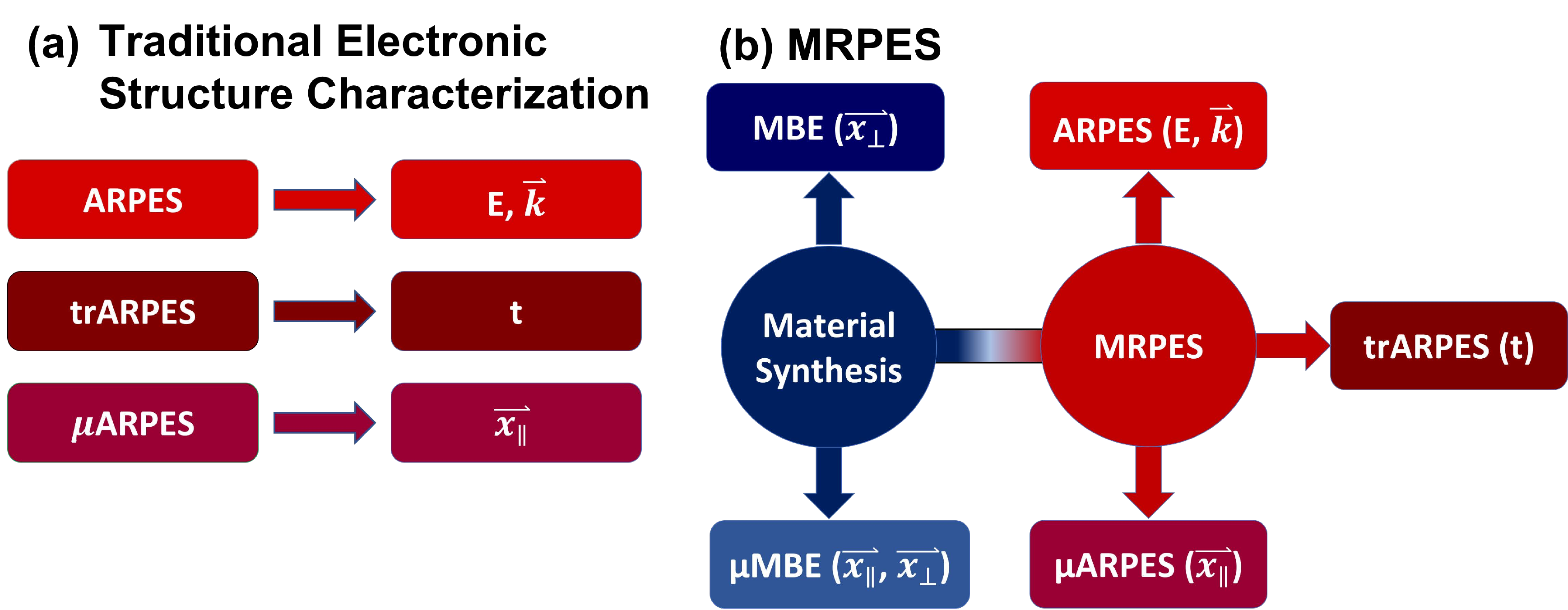}
    \caption{Development of a multi-resolution photoemission spectroscopy (MRPES) platform. (a) Typical methods of traditional electronic structure characterizations in which ARPES, trARPES, and $\mu$ARPES are separately employed to study different properties of a material. (b) The MRPES platform combines all elements of traditional electronic structure characterizations in connection with a customized molecular beam epitaxy (MBE) setup incorporating shadow masks.}
    \label{fig:MASTER}
\end{figure}

It is thus demanded to integrate multiple modalities in one ARPES setup. This integration brings fundamental challenges if using a single light source. For instance, the energy and temporal widths of light pulses are conjugate quantities dictated by the fundamental uncertainty principle~\cite{Gauthier2020}. A sub-100-fs time resolution in trARPES is concomitant with an energy resolution $> 18$~meV in static ARPES. Moreover, micron or sub-micron-sized beam spots in $\mu$ARPES or nanoARPES can lead to a significant space-charging effect which is detrimental in high-energy-resolution applications. This problem is particularly severe for photon energies > 10 eV.~\cite{Rotenberg2014, He2016}

In this report, we present a new integrated platform for multi-resolution photoemission spectroscopy (MRPES). This MRPES setup integrates a Helium discharge lamp, a narrow bandwidth 6 eV laser, and a tunable ultrafast laser, which effectively combines static ARPES, trARPES, and $\mu$ARPES. This setup provides a first-time all-in-one solution for multi-modal photoemission spectroscopy (Fig. 1(b)). From calibration experiments on Bi$_2$Se$_3$ and MnBi$_2$Te$_4$ we demonstrate an energy resolution $< 4$~meV in static laser-based ARPES, a time resolution of $35$~fs in trARPES, and a spatial resolution of $\sim 10$~$\mu$m in $\mu$ARPES. Remarkably, our time resolution sets a new record for trARPES setups with probe pulses generated by solid-state nonlinear crystals. In connection with a molecular beam epitaxy (MBE) system incorporating shadow masks, our setup enables a holistic probing of quantum materials which are engineered both in-plane and out-of-plane. The MRPES setup combines complementary capabilities of low and high photon energies, of energy and time resolutions, and of spatial and momentum probes. We demonstrate the performance of this novel setup using FeSe/SrTiO$_3$ thin film superconductors and MnBi$_4$Te$_7$ magnetic topological insulators.

In the remainder of this report, we will provide an overview of the system layout (Section \ref{sec:ExperimentalSetup}), demonstrate its energy, momentum, spatial, and temporal resolutions (Section \ref{sec:Calibration}), and showcase its versatility to control and probe materials such as FeSe/SrTiO$_3$ and MnBi$_4$Te$_7$ (Section \ref{sec:Applications}).

\section{Experimental Setup}
\label{sec:ExperimentalSetup}

\begin{figure}
    \centering
    \includegraphics[width = 3.5 in]{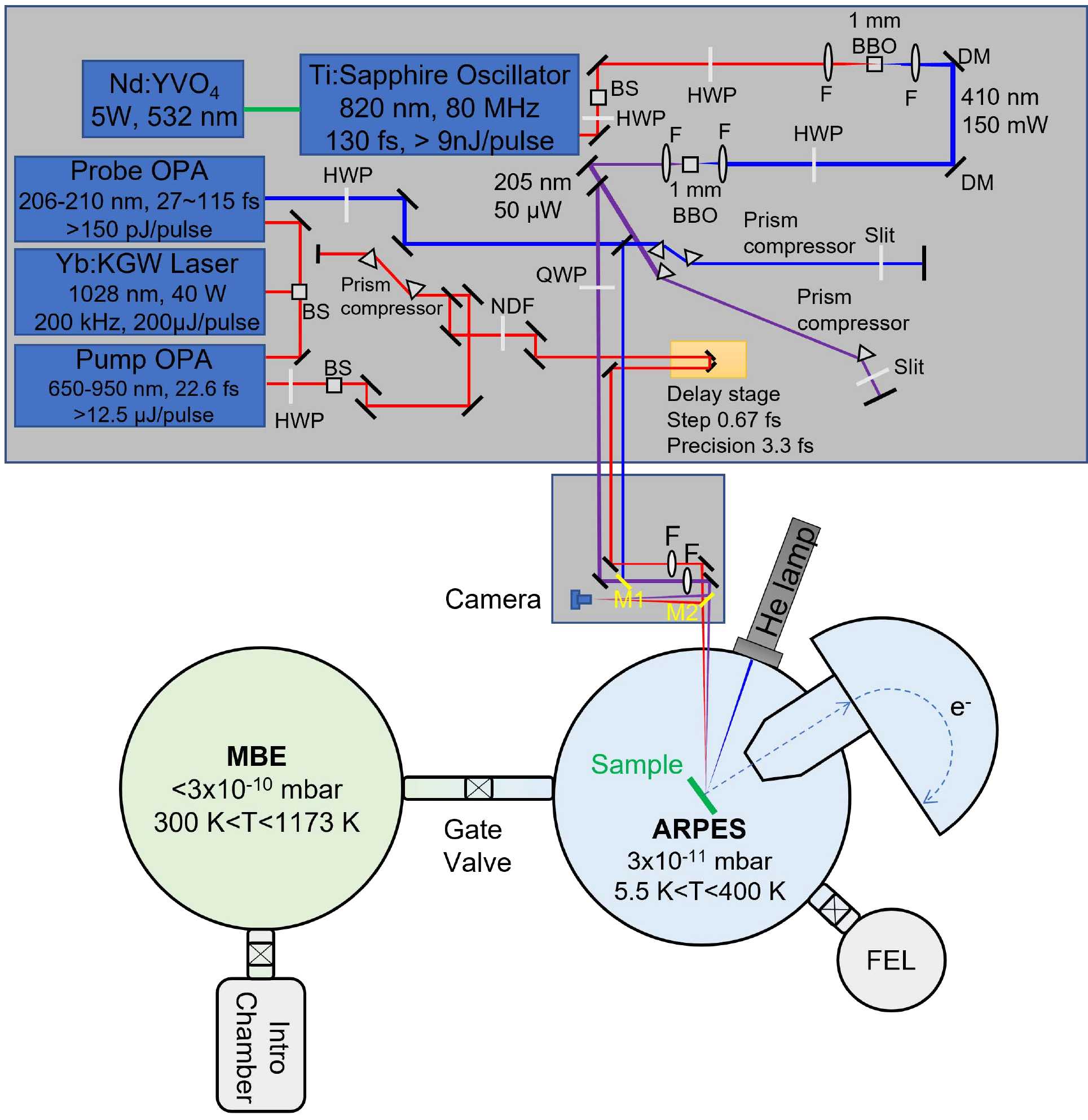}
    \caption{Schematic layout of the MRPES platform. HWP, QWP, BS, DM, F, NDF, and FEL stand for half waveplates, quarter waveplates, beam splitters, dichroic mirrors, focusing lenses, neutral density filters, and fast-entrance loadlock, respectively.}
    \label{fig:Lab_Schematic}
\end{figure}

The MRPES platform is schematically illustrated in Fig.~\ref{fig:Lab_Schematic}. Multiple light sources are connected to the photoemission chamber. The first light source is a 6 eV beam operating at a repetition rate of 80 MHz. A mode locked Ti:Sapphire oscillator (Coherent MIRA Optima 900) is pumped by a $5$~W continuous wave seed laser (Coherent Verdi V5). The Ti:Sapphire oscillator is tunable between $780$ and $900$~nm, and is typically set at $820$~nm with $9$~nJ pulse energies. $820$~nm pulses pass through two stages of second harmonic generation using $\beta$-barium borate (BBO) crystals to yield the fourth harmonic (205 nm) used for photoemission. The fourth harmonic is separated by a prism-pair compressor. A quarter waveplate (QWP) is used to switch between linear and circular polarizations.

The second light source is optimized for trARPES and begins with a diode-pumped Yb:KGW laser (Light Conversion Carbide C3). The laser outputs $200$~$\mu$J, $250$~fs pulses with a central wavelength at $1028$~nm and a repetition rate of $200$~kHz. This beam is split by a polarizing beam splitter and sent into two non-collinear optical parametric amplifiers (OPAs). $190$~$\mu$J are sent into a pump OPA to generate a tunable \textit{signal} beam between 650 nm and 950 nm, and is typically set at 800 nm. The output pulses have a pulse duration of $22.6$~fs and a pulse energy $> 12.5$~$\mu$J. The pump beam is directed to a delay stage which varies the beam path length by up to 30 cm (1 ns).

The remaining $10$~$\mu$J from Carbide are sent to the probe OPA where $0.9$~$\mu$J of \textit{signal} at $824$~nm is generated. $0.1$~$\mu$J of the second harmonic ($412$~nm) and $> 150$~pJ of the fourth harmonic ($206$~nm) are obtained using BBO crystals built into the probe OPA. The pulse duration of the $206$~nm beam varies between $27$ and $115$~fs, which is tuned by the thicknesses of the BBO crystals \cite{Gauthier2020}. A prism-pair compressor compensates for the group velocity dispersion (GVD) and controls the bandwidth of the $206$~nm beam. The addition of a remountable mirror M1 switches between the $6$~eV beams from the Ti:Sapphire oscillator and from the probe OPA. The remountable mirror M2 can be used to route both the pump and probe beams to a profiling camera where the spatial overlap is established.

The final light source is a Helium discharge lamp (Scienta VUV5000) which is integrated into the MRPES system \textit{in vacuo}. A grating-based monochromator can switch between the He $1\alpha$ (21.2 eV), He $1\beta$ (23.1 eV), or He $2\alpha$ (40.8 eV) lines allowing access to full Brillouin zones of most materials, which cannot be accomplished using 6 eV beams.

Integrating three light sources in the MRPES system realizes complementary functionalities. For instance, high photon energies from the Helium discharge lamp reveal the entirety of a Brillouin zone, while the Ti:Sapphire-based 6 eV laser provides a surgical, ultrahigh energy-resolution probe of the band structure near the zone center. With an 80 MHz repetition rate, the Ti:Sapphire-based 6 eV laser minimizes the space-charging effect~\cite{Corder2018}, and is an ideal source to optimize energy and spatial resolutions. The 200 kHz Yb:KGW system in combination with the non-collinear OPA's is an ideal light source for ultrafast trARPES measurements.

The heart of the MRPES system is a hemispherical electron analyzer (Scienta DA30-L).  The featured deflection voltage enables measurements of electrons emitted up to 15 deg with respect to the surface normal, which leads to convenient band mapping in a limited momentum range without the need to rotate samples. This is particularly important for precise measurements of micron-scale domains. Samples are loaded into a 6-axis manipulator reaching $5.5$~K with liquid Helium. The measurement chamber pressure is at $6\times 10^{-11}$~mbar when the manipulator is at room temperature, and drops to $3\times 10^{-11}$~mbar at the base temperature. This excellent vacuum condition leads to an extended sample lifetime $> 3$~days as characterized by photoemission experiments.

The MRPES system is connected under ultrahigh vacuum (UHV) to the MBE growth chamber which has a base pressure of 3 $\times$ 10$^{-10}$ mbar. Substrates can be loaded into MBE separately from the ARPES system through an introduction chamber. The MBE system features up to seven effusion cells, as well as a manipulator providing radiative heating up to $1100^{\circ}$~C or direct current heating up to $\sim 1500^{\circ}$~C. Notably, we implement laser-cut stainless-steel shadow masks to perform patterned growth. This shadow mask technique allows us to print micron-scale patterns without the need of exposing thin films to ambient or harsh environments for lithographic processing~\cite{Zhou2003, Tsioutsios2020} and is thus termed \emph{$\mu$MBE}. The printing resolution is currently limited by the laser cutting precision $\sim 5$~$\mu$m. In combination with $\mu$ARPES, our system enables a new modality to perform microscopic synthesis and characterization on novel quantum materials. A proof-of-principle experiment is shown in Fig.~\ref{fig:FeSe}.

\section{MRPES Calibration}
\label{sec:Calibration}

\begin{figure}
    \centering
    \includegraphics[width = 3.5 in]{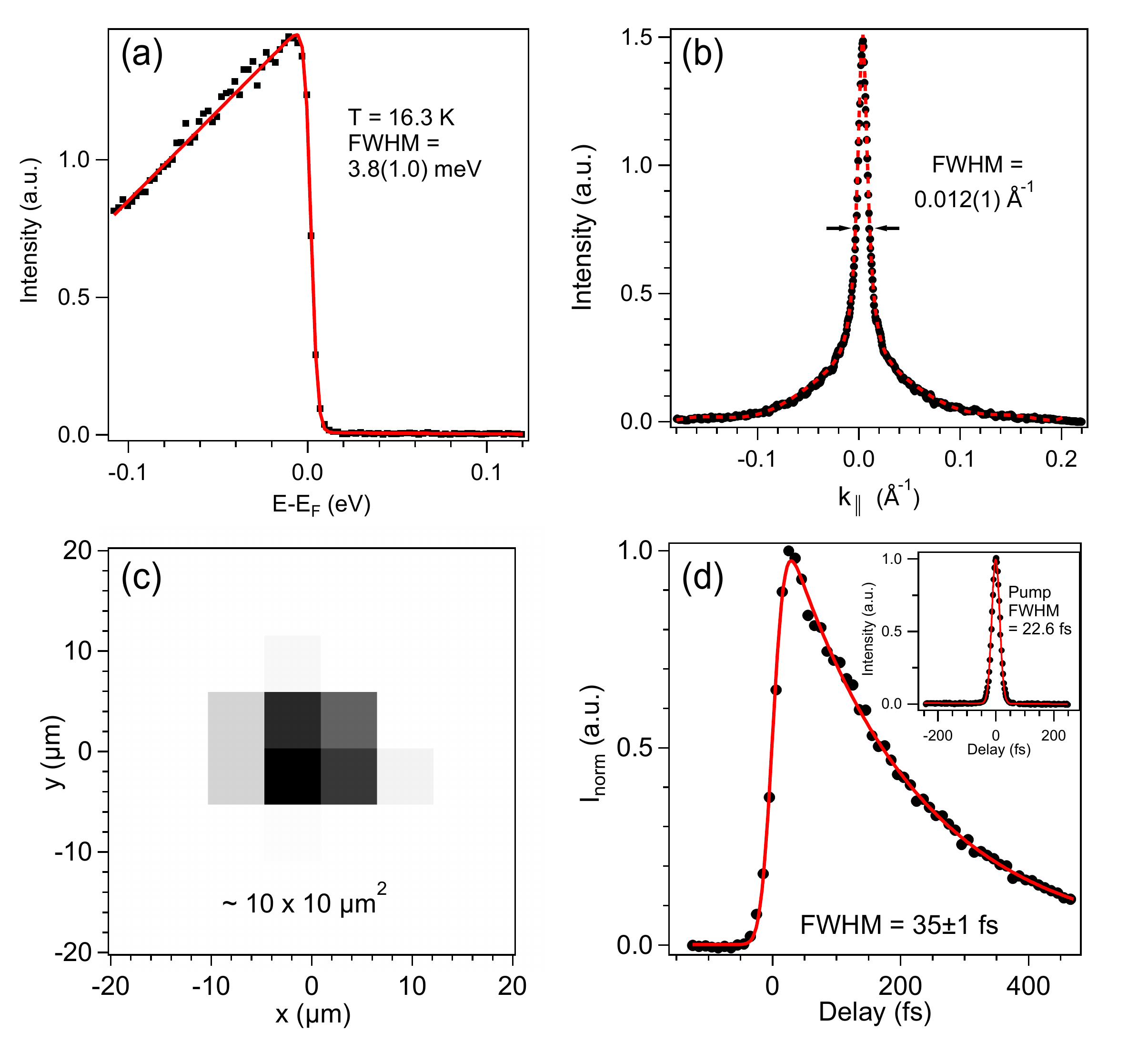}
    \caption{System calibrations. (a) Energy distribution curve (EDC) of Bi$_2$Se$_3$ taken using the Ti:Sappphire beam path (solid squares) and fit with a modified Fermi-Dirac function (solid line). (b) Momentum distribution curve (MDC) taken at the Dirac point of MnBi$_2$Te$_4$ using the Ti:Sapphire beam path (solid circles) and fit to a Lorentzian with a polynomial background term (dashed line). (c) Profile of the Ti:Sapphire based 6 eV beam. The pixel size of the camera is $5.6$~$\mu$m. (d) Pump-probe cross-correlation taken at 1~eV above the Fermi level of Bi$_2$Se$_3$ using the trARPES beam path (solid circles) fit to an exponential decay convolved with a Gaussian function (solid line). The inset illustrates the auto-correlation of the pump pulses, demonstrating a $22.6$~fs FWHM at $800$~nm.}
    \label{fig:Calibration}
\end{figure}

Our best energy resolution is realized by the 80 MHz Ti:Sapphire based beamline. We characterize the overall energy resolution using an exemplary measurement on a cleaved single-crystal Bi$_2$Se$_3$ at $16.3$~K. Figure \ref{fig:Calibration}(a) shows an energy distribution curve (EDC) taken through the $\overline{\Gamma}$ point near the Fermi level (E$_\textrm{F}$). The EDC was fit to a a modified Fermi-Dirac (FD) function, where the original FD distribution is multiplied with a linear density-of-states function and convolved with a Gaussian resolution function. The energy resolution, $3.8\pm 1.0$~meV, is close to the theoretical best overall energy resolution of $3.4$~meV. This theoretical estimate is based on the analyzer resolution of $2$~meV and the $6$~eV laser bandwidth of $2.7$~meV.~\cite{Gauthier2020}

While it is difficult to characterize our momentum resolution, we present an upper limit by showing a momentum distribution curve (MDC) taken at the Dirac point of an antiferromagnetic topological insulator MnBi$_2$Te$_4$ (Fig. \ref{fig:Calibration}(b)). The Ti:Sapphire oscillator is again used for this characterization. The full width at half maximum (FWHM) is determined to be $0.012\pm 0.001$~\AA$^{-1}$ by fitting the MDC to a Lorentzian function. We emphasize that this width is limited by material physics and only provides an upper limit for the momentum resolution.

The spatial resolution of the $\mu$ARPES module is characterized by imaging the focused beam profile. In a typical setup, the beam is first expanded to a diameter of $\sim 4$~mm, and then focused by a lens with a focal length of $250$~mm. The theoretical diffraction limited FWHM at the focal point is $9.6$~$\mu$m.~\cite{Demtroder2014} The high repetition rate~\cite{Corder2018} and low photon energy~\cite{Zhou2005} both lead to a minimal space charging effect. Figure \ref{fig:Calibration}(c) shows an image of the focused beam profile using an Imaging Source DMK 23U618 camera. Even though we cannot extract the exact FWHM’s of the beam profile due to the $5.6$~$\mu$m pixel size, we determine that the central four pixels occupy $60.2$\% of the total integrated intensity. Notably, for a Gaussian beam, the integrated intensity within the central region defined by FWHM’s occupies $57.9$\% of the total integrated intensity. Hence we estimate the beam size to be $\sim 10\times10$~$\mu$m$^2$. Importantly, the projected horizontal beam waist will be enlarged by a factor of $1/\cos{\theta}$ where $\theta$ is the laser's angle of incidence.

\begin{figure*}[t!]
    \centering
    \includegraphics[width = 7 in]{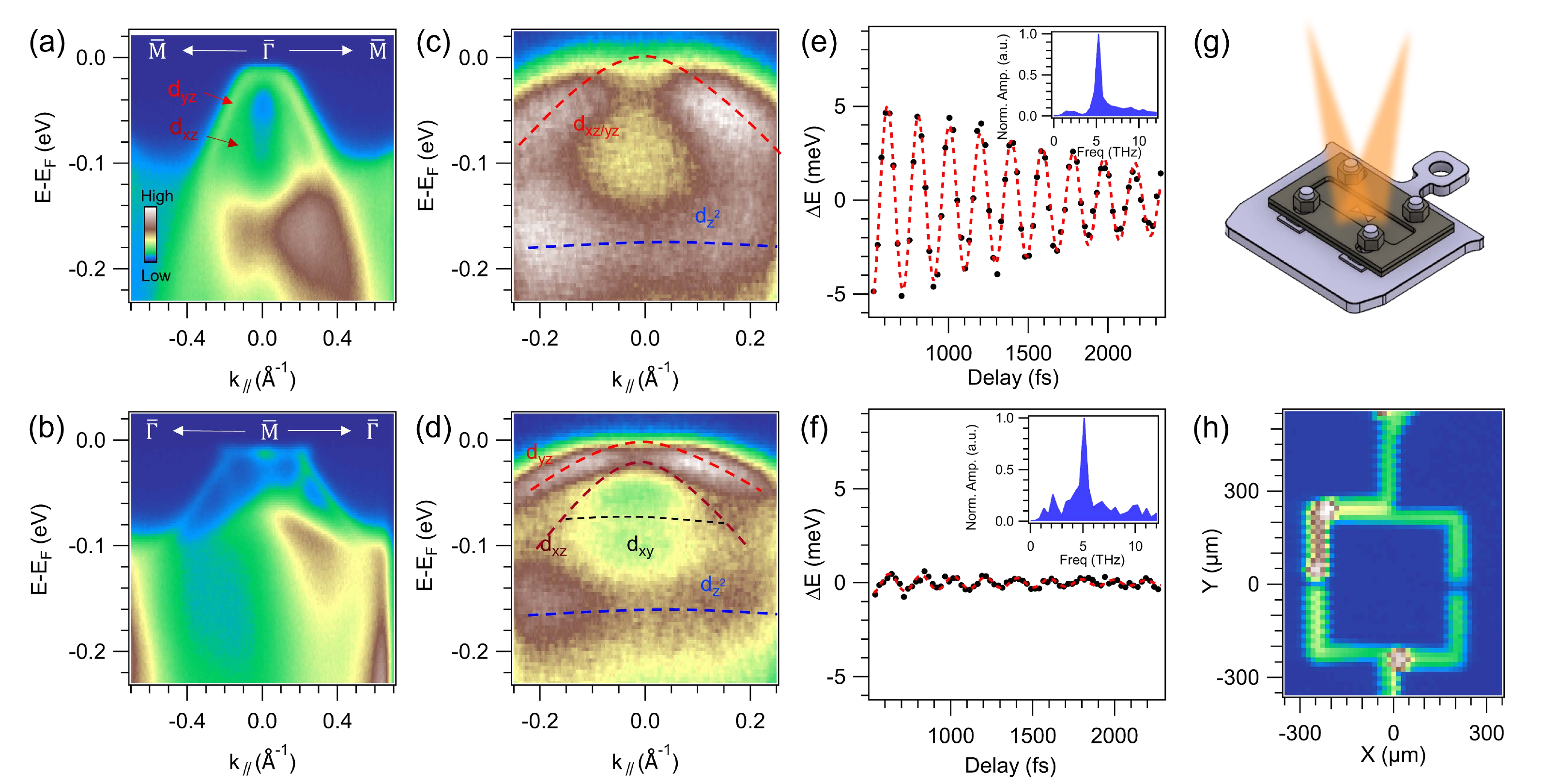}
    \caption{MRPES demonstrative experiments on FeSe/SrTiO$_{3}$. (a, b) Helium lamp ARPES spectra of 8 unit-cell thick FeSe/SrTiO$_{3}$ taken near the $\overline{\Gamma}$ and $\overline{M}$ points, respectively. (c, d) ARPES spectra near $\overline{\Gamma}$ taken using the trARPES module with (c) the high-time-resolution setup and (d) the high-energy-resolution setup. Dashed lines serve as guides for the bands. (e, f) Extracted energy oscillations of the $d_{xz/yz}$ band from the trARPES measurements corresponding to the data shown in panels (c) and (d), respectively. The band oscillations (solid circles) are fit to a cosine function modulated by an exponential decay (dashed line). The insets show the fast Fourier transforms (FFT) of the corresponding oscillations. (g) Schematic of $\mu$MBE growth using a shadow mask. (h) Spatial map of the $\mu$MBE-patterned FeSe square ring taken using the $\mu$ARPES module.}
    \label{fig:FeSe}
\end{figure*}

We demonstrate the time resolution of the trARPES module using the 200 kHz ultrafast pump-probe setup. Prism-pair compressors compensate for any broadening due to the group velocity dispersion (GVD) in both pump and probe pulses. We perform a trARPES experiment on a Bi$_2$Se$_3$ sample with a pump fluence of $200 \ \mu$J/cm$^2$. A cross-correlation is extracted from 1 eV above E$_\textrm{F}$ and is fitted to an exponential function convolved with a Gaussian resolution function (Figure \ref{fig:Calibration}(d)). The fitting yields a FWHM = $35 \pm 1$ fs, which is to our knowledge the best time resolution among all trARPES setups using solid-crystal-based frequency upconversion~\cite{Gauthier2020, Ishida2014, Yang2019, Sobota2012, Graf2011, Wang2012a, Perfetti2007}. This time resolution is obtained using a $150$~$\mu$m-thick BBO for second harmonic generation (SHG), and a $50$~$\mu$m-thick BBO for fourth harmonic generation (FHG). Since our pump pulse duration is $22.6$~fs (inset of Fig. \ref{fig:Calibration}(d)), the FWHM duration of the probe pulse is determined to be $27$~fs. By utilizing a $500$~$\mu$m-thick SHG BBO and a $100$~$\mu$m-thick FHG BBO, and retuning the probe compressor, we can also obtain an energy resolution of $17$~meV and a time resolution of $115$~fs. The high-time-resolution and high-energy-resolution setups allow us to reveal sub-50 fs dynamics and sub-20 meV energy features, respectively.

\section{Material Application}
\label{sec:Applications}
To demonstrate the multi-modalities of the MRPES system, we present studies on FeSe/SrTiO$_{3}$ thin film superconductors and MnBi$_{4}$Te$_{7}$ antiferromagnetic topological insulators. As demonstrated below, MRPES resolves fine features of FeSe/SrTiO$_{3}$ in both the energy and time domains, and disentangles termination-dependent topological properties of MnBi$_{4}$Te$_{7}$.

\subsection{FeSe}
FeSe/SrTiO$_{3}$ thin film superconductors~\cite{Wang2012,Lee2014,Song2019,Yang2015thick,Gerber2017,Suzuki2019} are model systems where the key electronic interactions occur at multiple energy scales. A strong coupling between the FeSe electrons and the $100$~meV SrTiO$_{3}$ phonons leads to a substantial boost of the superconducting gap in monolayer FeSe/SrTiO$_{3}$.~\cite{Lee2014,Song2019} Cooperative electron-electron and electron-phonon interactions yields an orders-of-magnitude enhancement to the coupling with the $22$~meV Se A$_{1g}$ mode~\cite{Yang2015thick, Gerber2017,Suzuki2019}. Resolving these different interactions on a single sample necessitates the usage of the MRPES setup. We grow an $8$-unit-cell thick ($8$-UC) FeSe thin film using the MBE module following recipes in the literature~\cite{Lee2014, Faeth2021} and transfer the sample \emph{in vacuo} to the MRPES module for multi-resolution measurements. 

Traditional ARPES measurements based on the Helium discharge lamp are first conducted to resolve the electronic band structure across the entire Brillouin zone. Photoemission spectra taken at $26$~K near the $\overline{\Gamma}$ and $\overline{M}$ points are shown in Fig.~\ref{fig:FeSe}(a) and \ref{fig:FeSe}(b), respectively. Two distinct hole-like bands are resolved near $\overline{\Gamma}$, which are attributed to the $d_{xz}$ and $d_{yz}$ orbitals. Near the $\overline{M}$ point a complex electronic band structure involving at least two electron-like and one hole-like pockets is resolved, which is due to the nematic splitting of all three $t_{2g}$ orbitals~\cite{Zhang2016, Yi2019}. These findings agree with those in the literature~\cite{Zhang2016, Yi2019, tan2013}, and establish the high quality of our thin films.

\begin{figure*}[ht!]
    \centering
    \includegraphics[width = 7 in]{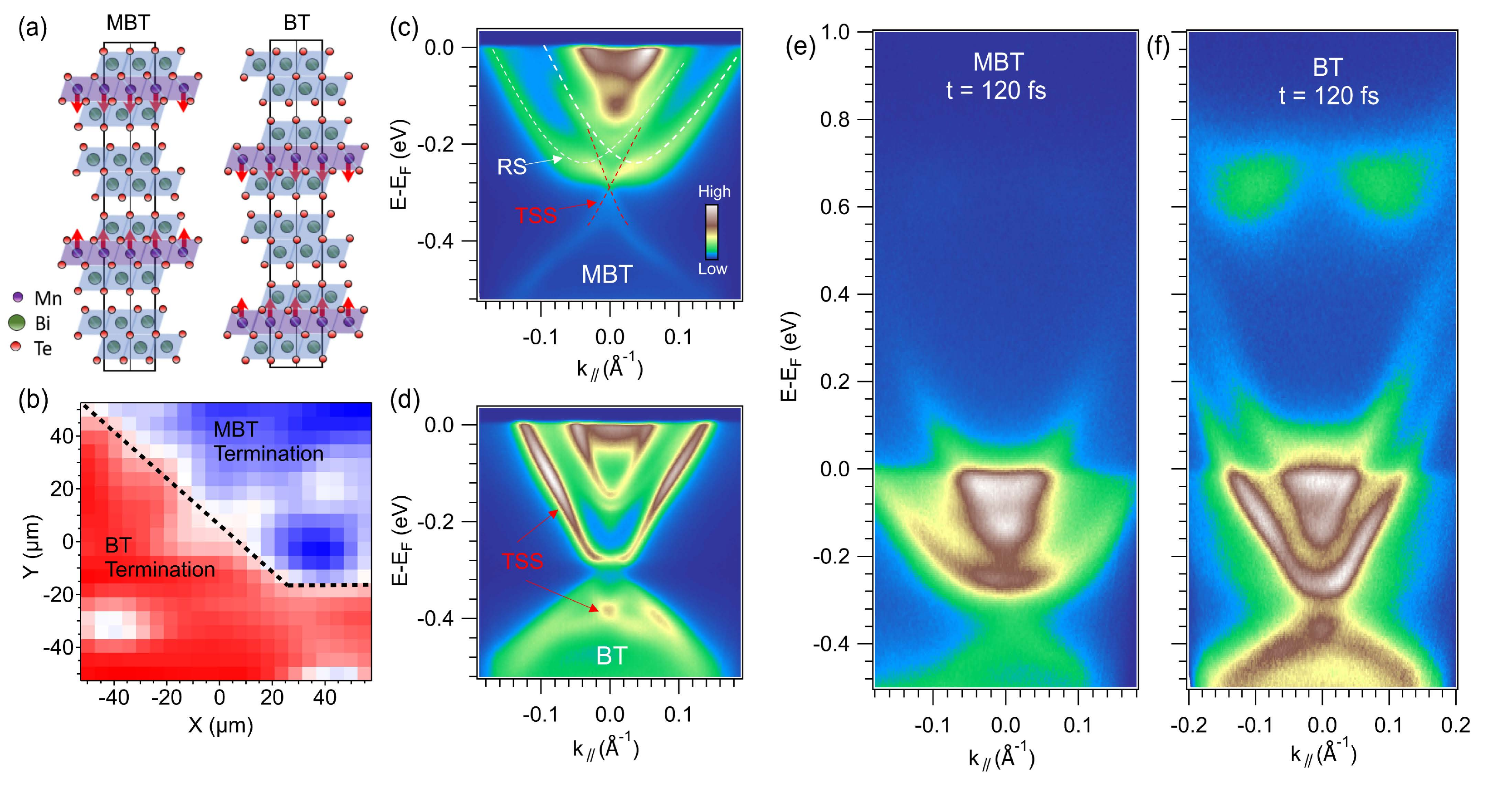}
    \caption{MRPES demonstative experiments on MnBi$_4$Te$_7$. (a) MnBi$_4$Te$_7$ crystal structures corresponding to the two possible terminations: MnBi$_2$Te$_4$ (MBT) and Bi$_2$Te$_3$ (BT). (b) Spatial mapping of an MnBi$_4$Te$_7$ cleaved surface taken with the $\mu$ARPES module. Red denotes the BT termination while blue denotes the MBT termination. (c, d) $\mu$ARPES spectra near $\overline{\Gamma}$ on (c) the MBT termination and (d) the BT termination. Dashed lines indicate the topological surface states (TSS) and Rashba states (RS). (e, f) Spectra taken with the trARPES module at 120 fs near $\overline{\Gamma}$ showing the unoccupied band structures of (e) the MBT termination and (f) the BT termination.}
    \label{fig:MBT}
\end{figure*}

To understand the key electronic interactions, we perform trARPES using both the high-time-resolution setup where the time and energy resolutions are $50$~fs and $40$~meV, as well as a high-energy-resolution setup where the time and energy resolutions are $115$~fs and $17$~meV. The pump fluence is maintained at $0.31$~mJ/cm$^{2}$ and the manipulator temperature is kept at $12$~K. Notably, the $d_{xz}$ and $d_{yz}$ orbitals cannot be separately resolved in trARPES due to finite energy resolutions, and we denote the overlapping spectral feature as the $d_{xz/yz}$ band. 

We extract the time-dependent band shift of the $d_{xz/yz}$ band by fitting the EDCs to a 2-band model~\cite{Gerber2017}. The non-oscillatory component of the band shift dynamics is approximated using a 7th-order polynomial. We obtain the oscillation amplitude by fitting the oscillatory component to a cosine function modulated by an exponential decay. For the high-time-resolution setup (Fig.~\ref{fig:FeSe}(c)), a prominent oscillation near $5$~THz is observed with an amplitude of 7.95 meV at time zero, which is attributed to the Se A$_{1g}$ mode~\cite{Yang2015thick, Gerber2017, Suzuki2019}. In contrast, using the high-energy-resolution setup (Fig.~\ref{fig:FeSe}(d)) we resolve oscillations with an amplitude of only 0.73 meV at time zero. The reduction of the oscillation amplitude in the latter measurement results from the convolution with the 115 fs time resolution \cite{Gerber2017}. This difference clearly demonstrates the advantage of the high-time-resolution setup in resolving fine features in the time domain.

The benefits of the high-energy-resolution setup can be seen in the energy domain (Fig.~\ref{fig:FeSe}(d)). With a $17$~meV energy resolution, trARPES not only resolves overall sharper band dispersions, but more importantly allows the identification of the weak $d_{xy}$ band. This band is of particular interest as it undergoes an orbital selective Mott transition at high temperatures \cite{huang2021}.

In addition to characterizing materials, the MRPES platform allows for micro-fabrication and micro-characterization based on the $\mu$MBE and $\mu$ARPES modules. In $\mu$MBE we grow a patterned 4-UC FeSe/SrTiO$_{3}$ film using a shadow mask (Fig.~\ref{fig:FeSe}(g)). A $5$~$\mu$m thick, stainless steel shadow mask is sandwiched between two Tantalum spacers and placed at 0.4 mm over an annealed SrTiO$_{3}$ substrate. The substrate-mask system is loaded into the MBE chamber and exposed to molecular beams. As a system test, we select a mask that features a square ring pattern with a 50~$\mu$m linewidth and a 20~$\mu$m gap between the upper and lower halves. The device quality is characterized using the $\mu$ARPES module. The $\mu$ARPES mapping based on overall photoemission counts on the detector (Fig.~\ref{fig:FeSe}(h)) reveals the targeted $50$~$\mu$m linewidth and $20$~$\mu$m gaps. Higher intensities on the left side are most likely due to the alignment between effusion cells and the shadow mask. A thinner shadow mask possibly made of Si$_{3}$N$_{4}$ and a shorter mask-substrate distance will improve the printing resolution down to $\sim 100$~nm.~\cite{Zhou2003, Tsioutsios2020, Grevin2011}

\subsection{MnBi$_4$Te$_7$}

The need for a multi-resolution approach to photoemission characterization can also be seen in MnBi$_4$Te$_7$. MnBi$_4$Te$_7$ is an antiferromagnetic topological insulator which consists of alternating  septuple  layers  of MnBi$_2$Te$_4$ (MBT) and quintuple layers of Bi$_2$Te$_3$ (BT)~\cite{Hu2020, Vidal2021, Wu2020, Hu2020a}. As such, there are two possible terminations when a MnBi$_{4}$Te$_{7}$ sample is cleaved (Fig.~\ref{fig:MBT}(a)). Domains of each termination are on the order of tens of $\mu$m with sharp boundaries, making it crucial to employ a microscopic beam size to ensure probing of single domains.

We first perform a spatial map of MnBi$_4$Te$_7$ at room temperature (Fig.~\ref{fig:MBT}(b)). The spatial map shows a clear distinction between adjacent MBT and BT domains. Measurements at $15$~K at selected MBT and BT locations provide a detailed view of the band structures (Fig. \ref{fig:MBT}(c) and \ref{fig:MBT}(d)). While the focus of this work is on the technical capabilities of MRPES, we note that the interpretations of various spectral features from MnBi$_{4}$Te$_{7}$ are currently under debates~\cite{Hu2020, Vidal2021, Ma2020}. Based on our previous studies~\cite{Yan2021, Yan2021a}, we attribute the large electron-like pocket on the MBT termination to a pair of Rashba-split states. On the BT termination, the topological surface state (TSS) hybridizes with the valence bands, resulting in the Dirac point located near -0.38~eV.

We also perform trARPES to reveal the unoccupied band structure. ARPES spectra near $\overline{\Gamma}$ are obtained at 120 fs on both terminations (Fig. \ref{fig:MBT}(e) and \ref{fig:MBT}(f)) using the high-energy-resolution ultrafast setup. The unoccupied parts of the conduction bands and TSS's are resolved up to $0.4$~eV above $E_{\textrm{F}}$ for both terminations. On the BT termination, additional features near $0.7$~eV are resolved. These high-energy features are results of direct optical transitions, and serve as an electron reservoir to fill the low-energy states~\cite{Sobota2012}. Obtaining information about both the occupied and unoccupied band structures is crucial for a complete analysis of the material and showcases the necessity of employing a multi-resolution system.

\section{Conclusion}
We have presented a MRPES platform that integrates helium lamp ARPES, static laser ARPES, trARPES, $\mu$ARPES, MBE, and $\mu$MBE into one system. The demonstrated energy resolution of $< 4$~meV for static ARPES, time resolution of $\sim 35$~fs for trARPES, and spatial resolution of $\sim 10$~$\mu$m for $\mu$ARPES allow for a holistic probing of materials' electronic properties. Our time resolution is the fastest among all trARPES setups using solid-crystal frequency upconversion. The integration with MBE and $\mu$MBE further enables material engineering both in- and out-of-plane, and opens the door to \emph{in situ} device fabrication. The performance of the MRPES system is benchmarked by the studies on thin film superconductors FeSe/SrTiO$_{3}$ and antiferromagnetic topological insulators MnBi$_{4}$Te$_{7}$. 

We emphasize that MRPES is not merely a combination of different ARPES modules. The simultaneous probing into energy, momentum, space and time, with uncompromised resolutions in each domain, allows us to fully reveal the complex physics in novel quantum materials. The adaptability and level of control unlock the ability to create and understand quantum devices such as superconducting qubits and superconducting quantum interference devices (SQUIDs) which transcends the traditional ARPES tool into a multi-modality quantum device characterization instrument. 

\begin{acknowledgements}
This work is partially supported by NSF grant \#2019131.  E.G.  acknowledges support by the UChicago Dean's Scholars program.  R.F.  acknowledges support by the UChicago Jeff Metcalf program.  The MnBi$_{4}$Te$_{7}$ sample preparation was supported by the National Science Foundation through the Penn State 2D Crystal Consortium-Materials Innovation Platform (2DCC-MIP) under NSF cooperative agreement DMR-1539916.  We would like to thank Jonathan Sobota,  Patrick Kirchmann,  Hadas Soifer,  Alexandre Gauthier,  Brendan Faeth, Kyle Shen, Darrell Schlom, and David Awschalom for helpful discussions.

The authors have no conflicts to disclose.
\end{acknowledgements}

\section*{Data Availability}
The data that support the findings of this study are available from the corresponding author upon reasonable request.

\bibliography{Citations.bib}

\begin{thebibliography}{51}%
\makeatletter
\providecommand \@ifxundefined [1]{%
 \@ifx{#1\undefined}
}%
\providecommand \@ifnum [1]{%
 \ifnum #1\expandafter \@firstoftwo
 \else \expandafter \@secondoftwo
 \fi
}%
\providecommand \@ifx [1]{%
 \ifx #1\expandafter \@firstoftwo
 \else \expandafter \@secondoftwo
 \fi
}%
\providecommand \natexlab [1]{#1}%
\providecommand \enquote  [1]{``#1''}%
\providecommand \bibnamefont  [1]{#1}%
\providecommand \bibfnamefont [1]{#1}%
\providecommand \citenamefont [1]{#1}%
\providecommand \href@noop [0]{\@secondoftwo}%
\providecommand \href [0]{\begingroup \@sanitize@url \@href}%
\providecommand \@href[1]{\@@startlink{#1}\@@href}%
\providecommand \@@href[1]{\endgroup#1\@@endlink}%
\providecommand \@sanitize@url [0]{\catcode `\\12\catcode `\$12\catcode
  `\&12\catcode `\#12\catcode `\^12\catcode `\_12\catcode `\%12\relax}%
\providecommand \@@startlink[1]{}%
\providecommand \@@endlink[0]{}%
\providecommand \url  [0]{\begingroup\@sanitize@url \@url }%
\providecommand \@url [1]{\endgroup\@href {#1}{\urlprefix }}%
\providecommand \urlprefix  [0]{URL }%
\providecommand \Eprint [0]{\href }%
\providecommand \doibase [0]{http://dx.doi.org/}%
\providecommand \selectlanguage [0]{\@gobble}%
\providecommand \bibinfo  [0]{\@secondoftwo}%
\providecommand \bibfield  [0]{\@secondoftwo}%
\providecommand \translation [1]{[#1]}%
\providecommand \BibitemOpen [0]{}%
\providecommand \bibitemStop [0]{}%
\providecommand \bibitemNoStop [0]{.\EOS\space}%
\providecommand \EOS [0]{\spacefactor3000\relax}%
\providecommand \BibitemShut  [1]{\csname bibitem#1\endcsname}%
\let\auto@bib@innerbib\@empty
\bibitem [{\citenamefont {Damascelli}, \citenamefont {Hussain},\ and\
  \citenamefont {Shen}(2003)}]{Damascelli2003}%
  \BibitemOpen
  \bibfield  {author} {\bibinfo {author} {\bibfnamefont {A.}~\bibnamefont
  {Damascelli}}, \bibinfo {author} {\bibfnamefont {Z.}~\bibnamefont {Hussain}},
  \ and\ \bibinfo {author} {\bibfnamefont {Z.-X.}\ \bibnamefont {Shen}},\
  }\bibfield  {title} {\enquote {\bibinfo {title} {Angle-resolved photoemission
  studies of the cuprate superconductors},}\ }\href {\doibase
  10.1103/RevModPhys.75.473} {\bibfield  {journal} {\bibinfo  {journal}
  {Reviews of Modern Physics}\ }\textbf {\bibinfo {volume} {75}},\ \bibinfo
  {pages} {473--541} (\bibinfo {year} {2003})}\BibitemShut {NoStop}%
\bibitem [{\citenamefont {Sobota}, \citenamefont {He},\ and\ \citenamefont
  {Shen}(2021)}]{Sobota2021}%
  \BibitemOpen
  \bibfield  {author} {\bibinfo {author} {\bibfnamefont {J.~A.}\ \bibnamefont
  {Sobota}}, \bibinfo {author} {\bibfnamefont {Y.}~\bibnamefont {He}}, \ and\
  \bibinfo {author} {\bibfnamefont {Z.-X.}\ \bibnamefont {Shen}},\ }\bibfield
  {title} {\enquote {\bibinfo {title} {Angle-resolved photoemission studies of
  quantum materials},}\ }\href {\doibase 10.1103/RevModPhys.93.025006}
  {\bibfield  {journal} {\bibinfo  {journal} {Reviews of Modern Physics}\
  }\textbf {\bibinfo {volume} {93}},\ \bibinfo {pages} {025006} (\bibinfo
  {year} {2021})}\BibitemShut {NoStop}%
\bibitem [{\citenamefont {Shen}\ \emph {et~al.}(1993)\citenamefont {Shen},
  \citenamefont {Dessau}, \citenamefont {Wells}, \citenamefont {King},
  \citenamefont {Spicer}, \citenamefont {Arko}, \citenamefont {Marshall},
  \citenamefont {Lombardo}, \citenamefont {Kapitulnik}, \citenamefont
  {Dickinson}, \citenamefont {Doniach}, \citenamefont {DiCarlo}, \citenamefont
  {Loeser},\ and\ \citenamefont {Park}}]{Shen1993}%
  \BibitemOpen
  \bibfield  {author} {\bibinfo {author} {\bibfnamefont {Z.-X.}\ \bibnamefont
  {Shen}}, \bibinfo {author} {\bibfnamefont {D.~S.}\ \bibnamefont {Dessau}},
  \bibinfo {author} {\bibfnamefont {B.~O.}\ \bibnamefont {Wells}}, \bibinfo
  {author} {\bibfnamefont {D.~M.}\ \bibnamefont {King}}, \bibinfo {author}
  {\bibfnamefont {W.~E.}\ \bibnamefont {Spicer}}, \bibinfo {author}
  {\bibfnamefont {A.~J.}\ \bibnamefont {Arko}}, \bibinfo {author}
  {\bibfnamefont {D.}~\bibnamefont {Marshall}}, \bibinfo {author}
  {\bibfnamefont {L.~W.}\ \bibnamefont {Lombardo}}, \bibinfo {author}
  {\bibfnamefont {A.}~\bibnamefont {Kapitulnik}}, \bibinfo {author}
  {\bibfnamefont {P.}~\bibnamefont {Dickinson}}, \bibinfo {author}
  {\bibfnamefont {S.}~\bibnamefont {Doniach}}, \bibinfo {author} {\bibfnamefont
  {J.}~\bibnamefont {DiCarlo}}, \bibinfo {author} {\bibfnamefont
  {T.}~\bibnamefont {Loeser}}, \ and\ \bibinfo {author} {\bibfnamefont {C.~H.}\
  \bibnamefont {Park}},\ }\bibfield  {title} {\enquote {\bibinfo {title}
  {Anomalously large gap anisotropy in the a-b plane of
  {Bi}$_2${Sr}$_2${Ca}{Cu}$_2${O}$_{8+\delta}$},}\ }\href {\doibase
  10.1103/PhysRevLett.70.1553} {\bibfield  {journal} {\bibinfo  {journal}
  {Physical Review Letters}\ }\textbf {\bibinfo {volume} {70}},\ \bibinfo
  {pages} {1553--1556} (\bibinfo {year} {1993})}\BibitemShut {NoStop}%
\bibitem [{\citenamefont {Vishik}\ \emph {et~al.}(2012)\citenamefont {Vishik},
  \citenamefont {Hashimoto}, \citenamefont {He}, \citenamefont {Lee},
  \citenamefont {Schmitt}, \citenamefont {Lu}, \citenamefont {Moore},
  \citenamefont {Zhang}, \citenamefont {Meevasana}, \citenamefont {Sasagawa},
  \citenamefont {Uchida}, \citenamefont {Fujita}, \citenamefont {Ishida},
  \citenamefont {Ishikado}, \citenamefont {Yoshida}, \citenamefont {Eisaki},
  \citenamefont {Hussain}, \citenamefont {Devereaux},\ and\ \citenamefont
  {Shen}}]{Vishik2012}%
  \BibitemOpen
  \bibfield  {author} {\bibinfo {author} {\bibfnamefont {I.~M.}\ \bibnamefont
  {Vishik}}, \bibinfo {author} {\bibfnamefont {M.}~\bibnamefont {Hashimoto}},
  \bibinfo {author} {\bibfnamefont {R.-H.}\ \bibnamefont {He}}, \bibinfo
  {author} {\bibfnamefont {W.-S.}\ \bibnamefont {Lee}}, \bibinfo {author}
  {\bibfnamefont {F.}~\bibnamefont {Schmitt}}, \bibinfo {author} {\bibfnamefont
  {D.}~\bibnamefont {Lu}}, \bibinfo {author} {\bibfnamefont {R.~G.}\
  \bibnamefont {Moore}}, \bibinfo {author} {\bibfnamefont {C.}~\bibnamefont
  {Zhang}}, \bibinfo {author} {\bibfnamefont {W.}~\bibnamefont {Meevasana}},
  \bibinfo {author} {\bibfnamefont {T.}~\bibnamefont {Sasagawa}}, \bibinfo
  {author} {\bibfnamefont {S.}~\bibnamefont {Uchida}}, \bibinfo {author}
  {\bibfnamefont {K.}~\bibnamefont {Fujita}}, \bibinfo {author} {\bibfnamefont
  {S.}~\bibnamefont {Ishida}}, \bibinfo {author} {\bibfnamefont
  {M.}~\bibnamefont {Ishikado}}, \bibinfo {author} {\bibfnamefont
  {Y.}~\bibnamefont {Yoshida}}, \bibinfo {author} {\bibfnamefont
  {H.}~\bibnamefont {Eisaki}}, \bibinfo {author} {\bibfnamefont
  {Z.}~\bibnamefont {Hussain}}, \bibinfo {author} {\bibfnamefont {T.~P.}\
  \bibnamefont {Devereaux}}, \ and\ \bibinfo {author} {\bibfnamefont {Z.-X.}\
  \bibnamefont {Shen}},\ }\bibfield  {title} {\enquote {\bibinfo {title} {Phase
  competition in trisected superconducting dome},}\ }\href {\doibase
  10.1073/pnas.1209471109} {\bibfield  {journal} {\bibinfo  {journal}
  {Proceedings of the National Academy of Sciences}\ }\textbf {\bibinfo
  {volume} {109}},\ \bibinfo {pages} {18332--18337} (\bibinfo {year}
  {2012})}\BibitemShut {NoStop}%
\bibitem [{\citenamefont {Hsieh}\ \emph {et~al.}(2008)\citenamefont {Hsieh},
  \citenamefont {Qian}, \citenamefont {Wray}, \citenamefont {Xia},
  \citenamefont {Hor}, \citenamefont {Cava},\ and\ \citenamefont
  {Hasan}}]{Hsieh2008}%
  \BibitemOpen
  \bibfield  {author} {\bibinfo {author} {\bibfnamefont {D.}~\bibnamefont
  {Hsieh}}, \bibinfo {author} {\bibfnamefont {D.}~\bibnamefont {Qian}},
  \bibinfo {author} {\bibfnamefont {L.}~\bibnamefont {Wray}}, \bibinfo {author}
  {\bibfnamefont {Y.}~\bibnamefont {Xia}}, \bibinfo {author} {\bibfnamefont
  {Y.~S.}\ \bibnamefont {Hor}}, \bibinfo {author} {\bibfnamefont {R.~J.}\
  \bibnamefont {Cava}}, \ and\ \bibinfo {author} {\bibfnamefont {M.~Z.}\
  \bibnamefont {Hasan}},\ }\bibfield  {title} {\enquote {\bibinfo {title} {A
  topological {Dirac} insulator in a quantum spin {Hall} phase},}\ }\href
  {\doibase 10.1038/nature06843} {\bibfield  {journal} {\bibinfo  {journal}
  {Nature}\ }\textbf {\bibinfo {volume} {452}},\ \bibinfo {pages} {970--974}
  (\bibinfo {year} {2008})}\BibitemShut {NoStop}%
\bibitem [{\citenamefont {Chen}\ \emph {et~al.}(2009)\citenamefont {Chen},
  \citenamefont {Analytis}, \citenamefont {Chu}, \citenamefont {Liu},
  \citenamefont {Mo}, \citenamefont {Qi}, \citenamefont {Zhang}, \citenamefont
  {Lu}, \citenamefont {Dai}, \citenamefont {Fang}, \citenamefont {Zhang},
  \citenamefont {Fisher}, \citenamefont {Hussain},\ and\ \citenamefont
  {Shen}}]{Chen2009}%
  \BibitemOpen
  \bibfield  {author} {\bibinfo {author} {\bibfnamefont {Y.~L.}\ \bibnamefont
  {Chen}}, \bibinfo {author} {\bibfnamefont {J.~G.}\ \bibnamefont {Analytis}},
  \bibinfo {author} {\bibfnamefont {J.-H.}\ \bibnamefont {Chu}}, \bibinfo
  {author} {\bibfnamefont {Z.~K.}\ \bibnamefont {Liu}}, \bibinfo {author}
  {\bibfnamefont {S.-K.}\ \bibnamefont {Mo}}, \bibinfo {author} {\bibfnamefont
  {X.~L.}\ \bibnamefont {Qi}}, \bibinfo {author} {\bibfnamefont {H.~J.}\
  \bibnamefont {Zhang}}, \bibinfo {author} {\bibfnamefont {D.~H.}\ \bibnamefont
  {Lu}}, \bibinfo {author} {\bibfnamefont {X.}~\bibnamefont {Dai}}, \bibinfo
  {author} {\bibfnamefont {Z.}~\bibnamefont {Fang}}, \bibinfo {author}
  {\bibfnamefont {S.~C.}\ \bibnamefont {Zhang}}, \bibinfo {author}
  {\bibfnamefont {I.~R.}\ \bibnamefont {Fisher}}, \bibinfo {author}
  {\bibfnamefont {Z.}~\bibnamefont {Hussain}}, \ and\ \bibinfo {author}
  {\bibfnamefont {Z.-X.}\ \bibnamefont {Shen}},\ }\bibfield  {title} {\enquote
  {\bibinfo {title} {Experimental realization of a three-dimensional
  topological insulator, {Bi$_2$}{Te$_3$}},}\ }\href
  {https://www.science.org/doi/abs/10.1126/science.1173034} {\bibfield
  {journal} {\bibinfo  {journal} {Science}\ } (\bibinfo {year}
  {2009})}\BibitemShut {NoStop}%
\bibitem [{\citenamefont {Zhang}\ \emph {et~al.}(2018)\citenamefont {Zhang},
  \citenamefont {Yaji}, \citenamefont {Hashimoto}, \citenamefont {Ota},
  \citenamefont {Kondo}, \citenamefont {Okazaki}, \citenamefont {Wang},
  \citenamefont {Wen}, \citenamefont {Gu}, \citenamefont {Ding},\ and\
  \citenamefont {Shin}}]{zhang2018}%
  \BibitemOpen
  \bibfield  {author} {\bibinfo {author} {\bibfnamefont {P.}~\bibnamefont
  {Zhang}}, \bibinfo {author} {\bibfnamefont {K.}~\bibnamefont {Yaji}},
  \bibinfo {author} {\bibfnamefont {T.}~\bibnamefont {Hashimoto}}, \bibinfo
  {author} {\bibfnamefont {Y.}~\bibnamefont {Ota}}, \bibinfo {author}
  {\bibfnamefont {T.}~\bibnamefont {Kondo}}, \bibinfo {author} {\bibfnamefont
  {K.}~\bibnamefont {Okazaki}}, \bibinfo {author} {\bibfnamefont
  {Z.}~\bibnamefont {Wang}}, \bibinfo {author} {\bibfnamefont {J.}~\bibnamefont
  {Wen}}, \bibinfo {author} {\bibfnamefont {G.~D.}\ \bibnamefont {Gu}},
  \bibinfo {author} {\bibfnamefont {H.}~\bibnamefont {Ding}}, \ and\ \bibinfo
  {author} {\bibfnamefont {S.}~\bibnamefont {Shin}},\ }\bibfield  {title}
  {\enquote {\bibinfo {title} {Observation of topological superconductivity on
  the surface of an iron-based superconductor},}\ }\href
  {https://www.science.org/doi/abs/10.1126/science.aan4596} {\bibfield
  {journal} {\bibinfo  {journal} {Science}\ } (\bibinfo {year}
  {2018})}\BibitemShut {NoStop}%
\bibitem [{\citenamefont {Liu}\ \emph {et~al.}(2012)\citenamefont {Liu},
  \citenamefont {Zhang}, \citenamefont {Mou}, \citenamefont {He}, \citenamefont
  {Ou}, \citenamefont {Wang}, \citenamefont {Li}, \citenamefont {Wang},
  \citenamefont {Zhao}, \citenamefont {He}, \citenamefont {Peng}, \citenamefont
  {Liu}, \citenamefont {Chen}, \citenamefont {Yu}, \citenamefont {Liu},
  \citenamefont {Dong}, \citenamefont {Zhang}, \citenamefont {Chen},
  \citenamefont {Xu}, \citenamefont {Hu}, \citenamefont {Chen}, \citenamefont
  {Ma}, \citenamefont {Xue},\ and\ \citenamefont {Zhou}}]{Liu2012}%
  \BibitemOpen
  \bibfield  {author} {\bibinfo {author} {\bibfnamefont {D.}~\bibnamefont
  {Liu}}, \bibinfo {author} {\bibfnamefont {W.}~\bibnamefont {Zhang}}, \bibinfo
  {author} {\bibfnamefont {D.}~\bibnamefont {Mou}}, \bibinfo {author}
  {\bibfnamefont {J.}~\bibnamefont {He}}, \bibinfo {author} {\bibfnamefont
  {Y.-B.}\ \bibnamefont {Ou}}, \bibinfo {author} {\bibfnamefont {Q.-Y.}\
  \bibnamefont {Wang}}, \bibinfo {author} {\bibfnamefont {Z.}~\bibnamefont
  {Li}}, \bibinfo {author} {\bibfnamefont {L.}~\bibnamefont {Wang}}, \bibinfo
  {author} {\bibfnamefont {L.}~\bibnamefont {Zhao}}, \bibinfo {author}
  {\bibfnamefont {S.}~\bibnamefont {He}}, \bibinfo {author} {\bibfnamefont
  {Y.}~\bibnamefont {Peng}}, \bibinfo {author} {\bibfnamefont {X.}~\bibnamefont
  {Liu}}, \bibinfo {author} {\bibfnamefont {C.}~\bibnamefont {Chen}}, \bibinfo
  {author} {\bibfnamefont {L.}~\bibnamefont {Yu}}, \bibinfo {author}
  {\bibfnamefont {G.}~\bibnamefont {Liu}}, \bibinfo {author} {\bibfnamefont
  {X.}~\bibnamefont {Dong}}, \bibinfo {author} {\bibfnamefont {J.}~\bibnamefont
  {Zhang}}, \bibinfo {author} {\bibfnamefont {C.}~\bibnamefont {Chen}},
  \bibinfo {author} {\bibfnamefont {Z.}~\bibnamefont {Xu}}, \bibinfo {author}
  {\bibfnamefont {J.}~\bibnamefont {Hu}}, \bibinfo {author} {\bibfnamefont
  {X.}~\bibnamefont {Chen}}, \bibinfo {author} {\bibfnamefont {X.}~\bibnamefont
  {Ma}}, \bibinfo {author} {\bibfnamefont {Q.}~\bibnamefont {Xue}}, \ and\
  \bibinfo {author} {\bibfnamefont {X.~J.}\ \bibnamefont {Zhou}},\ }\bibfield
  {title} {\enquote {\bibinfo {title} {Electronic origin of high-temperature
  superconductivity in single-layer {FeSe} superconductor},}\ }\href {\doibase
  10.1038/ncomms1946} {\bibfield  {journal} {\bibinfo  {journal} {Nature
  Communications}\ }\textbf {\bibinfo {volume} {3}},\ \bibinfo {pages} {931}
  (\bibinfo {year} {2012})}\BibitemShut {NoStop}%
\bibitem [{\citenamefont {Tan}\ \emph {et~al.}(2013)\citenamefont {Tan},
  \citenamefont {Zhang}, \citenamefont {Xia}, \citenamefont {Ye}, \citenamefont
  {Chen}, \citenamefont {Xie}, \citenamefont {Peng}, \citenamefont {Xu},
  \citenamefont {Fan}, \citenamefont {Xu}, \citenamefont {Jiang}, \citenamefont
  {Zhang}, \citenamefont {Lai}, \citenamefont {Xiang}, \citenamefont {Hu},
  \citenamefont {Xie},\ and\ \citenamefont {Feng}}]{tan2013}%
  \BibitemOpen
  \bibfield  {author} {\bibinfo {author} {\bibfnamefont {S.}~\bibnamefont
  {Tan}}, \bibinfo {author} {\bibfnamefont {Y.}~\bibnamefont {Zhang}}, \bibinfo
  {author} {\bibfnamefont {M.}~\bibnamefont {Xia}}, \bibinfo {author}
  {\bibfnamefont {Z.}~\bibnamefont {Ye}}, \bibinfo {author} {\bibfnamefont
  {F.}~\bibnamefont {Chen}}, \bibinfo {author} {\bibfnamefont {X.}~\bibnamefont
  {Xie}}, \bibinfo {author} {\bibfnamefont {R.}~\bibnamefont {Peng}}, \bibinfo
  {author} {\bibfnamefont {D.}~\bibnamefont {Xu}}, \bibinfo {author}
  {\bibfnamefont {Q.}~\bibnamefont {Fan}}, \bibinfo {author} {\bibfnamefont
  {H.}~\bibnamefont {Xu}}, \bibinfo {author} {\bibfnamefont {J.}~\bibnamefont
  {Jiang}}, \bibinfo {author} {\bibfnamefont {T.}~\bibnamefont {Zhang}},
  \bibinfo {author} {\bibfnamefont {X.}~\bibnamefont {Lai}}, \bibinfo {author}
  {\bibfnamefont {T.}~\bibnamefont {Xiang}}, \bibinfo {author} {\bibfnamefont
  {J.}~\bibnamefont {Hu}}, \bibinfo {author} {\bibfnamefont {B.}~\bibnamefont
  {Xie}}, \ and\ \bibinfo {author} {\bibfnamefont {D.}~\bibnamefont {Feng}},\
  }\bibfield  {title} {\enquote {\bibinfo {title} {Interface-induced
  superconductivity and strain-dependent spin density waves in
  {FeSe}/{SrTiO$_3$} thin films},}\ }\href {\doibase 10.1038/nmat3654}
  {\bibfield  {journal} {\bibinfo  {journal} {Nature Materials}\ }\textbf
  {\bibinfo {volume} {12}},\ \bibinfo {pages} {634--640} (\bibinfo {year}
  {2013})}\BibitemShut {NoStop}%
\bibitem [{\citenamefont {He}\ \emph {et~al.}(2013)\citenamefont {He},
  \citenamefont {He}, \citenamefont {Zhang}, \citenamefont {Zhao},
  \citenamefont {Liu}, \citenamefont {Liu}, \citenamefont {Mou}, \citenamefont
  {Ou}, \citenamefont {Wang}, \citenamefont {Li}, \citenamefont {Wang},
  \citenamefont {Peng}, \citenamefont {Liu}, \citenamefont {Chen},
  \citenamefont {Yu}, \citenamefont {Liu}, \citenamefont {Dong}, \citenamefont
  {Zhang}, \citenamefont {Chen}, \citenamefont {Xu}, \citenamefont {Chen},
  \citenamefont {Ma}, \citenamefont {Xue},\ and\ \citenamefont
  {Zhou}}]{he2013}%
  \BibitemOpen
  \bibfield  {author} {\bibinfo {author} {\bibfnamefont {S.}~\bibnamefont
  {He}}, \bibinfo {author} {\bibfnamefont {J.}~\bibnamefont {He}}, \bibinfo
  {author} {\bibfnamefont {W.}~\bibnamefont {Zhang}}, \bibinfo {author}
  {\bibfnamefont {L.}~\bibnamefont {Zhao}}, \bibinfo {author} {\bibfnamefont
  {D.}~\bibnamefont {Liu}}, \bibinfo {author} {\bibfnamefont {X.}~\bibnamefont
  {Liu}}, \bibinfo {author} {\bibfnamefont {D.}~\bibnamefont {Mou}}, \bibinfo
  {author} {\bibfnamefont {Y.-B.}\ \bibnamefont {Ou}}, \bibinfo {author}
  {\bibfnamefont {Q.-Y.}\ \bibnamefont {Wang}}, \bibinfo {author}
  {\bibfnamefont {Z.}~\bibnamefont {Li}}, \bibinfo {author} {\bibfnamefont
  {L.}~\bibnamefont {Wang}}, \bibinfo {author} {\bibfnamefont {Y.}~\bibnamefont
  {Peng}}, \bibinfo {author} {\bibfnamefont {Y.}~\bibnamefont {Liu}}, \bibinfo
  {author} {\bibfnamefont {C.}~\bibnamefont {Chen}}, \bibinfo {author}
  {\bibfnamefont {L.}~\bibnamefont {Yu}}, \bibinfo {author} {\bibfnamefont
  {G.}~\bibnamefont {Liu}}, \bibinfo {author} {\bibfnamefont {X.}~\bibnamefont
  {Dong}}, \bibinfo {author} {\bibfnamefont {J.}~\bibnamefont {Zhang}},
  \bibinfo {author} {\bibfnamefont {C.}~\bibnamefont {Chen}}, \bibinfo {author}
  {\bibfnamefont {Z.}~\bibnamefont {Xu}}, \bibinfo {author} {\bibfnamefont
  {X.}~\bibnamefont {Chen}}, \bibinfo {author} {\bibfnamefont {X.}~\bibnamefont
  {Ma}}, \bibinfo {author} {\bibfnamefont {Q.}~\bibnamefont {Xue}}, \ and\
  \bibinfo {author} {\bibfnamefont {X.~J.}\ \bibnamefont {Zhou}},\ }\bibfield
  {title} {\enquote {\bibinfo {title} {Phase diagram and electronic indication
  of high-temperature superconductivity at 65 {K} in single-layer {FeSe}
  films},}\ }\href {\doibase 10.1038/nmat3648} {\bibfield  {journal} {\bibinfo
  {journal} {Nature Materials}\ }\textbf {\bibinfo {volume} {12}},\ \bibinfo
  {pages} {605--610} (\bibinfo {year} {2013})}\BibitemShut {NoStop}%
\bibitem [{\citenamefont {Iwasawa}\ \emph {et~al.}(2017)\citenamefont
  {Iwasawa}, \citenamefont {Schwier}, \citenamefont {Arita}, \citenamefont
  {Ino}, \citenamefont {Namatame}, \citenamefont {Taniguchi}, \citenamefont
  {Aiura},\ and\ \citenamefont {Shimada}}]{Iwasawa2017}%
  \BibitemOpen
  \bibfield  {author} {\bibinfo {author} {\bibfnamefont {H.}~\bibnamefont
  {Iwasawa}}, \bibinfo {author} {\bibfnamefont {E.~F.}\ \bibnamefont
  {Schwier}}, \bibinfo {author} {\bibfnamefont {M.}~\bibnamefont {Arita}},
  \bibinfo {author} {\bibfnamefont {A.}~\bibnamefont {Ino}}, \bibinfo {author}
  {\bibfnamefont {H.}~\bibnamefont {Namatame}}, \bibinfo {author}
  {\bibfnamefont {M.}~\bibnamefont {Taniguchi}}, \bibinfo {author}
  {\bibfnamefont {Y.}~\bibnamefont {Aiura}}, \ and\ \bibinfo {author}
  {\bibfnamefont {K.}~\bibnamefont {Shimada}},\ }\bibfield  {title} {\enquote
  {\bibinfo {title} {Development of laser-based scanning µ-{ARPES} system with
  ultimate energy and momentum resolutions},}\ }\href {\doibase
  10.1016/j.ultramic.2017.06.016} {\bibfield  {journal} {\bibinfo  {journal}
  {Ultramicroscopy}\ }\textbf {\bibinfo {volume} {182}},\ \bibinfo {pages}
  {85--91} (\bibinfo {year} {2017})}\BibitemShut {NoStop}%
\bibitem [{\citenamefont {Barbo}\ \emph {et~al.}(2000)\citenamefont {Barbo},
  \citenamefont {Bertolo}, \citenamefont {Bianco}, \citenamefont {Cautero},
  \citenamefont {Fontana}, \citenamefont {Johal}, \citenamefont {La~Rosa},
  \citenamefont {Margaritondo},\ and\ \citenamefont {Kaznacheyev}}]{Barbo2000}%
  \BibitemOpen
  \bibfield  {author} {\bibinfo {author} {\bibfnamefont {F.}~\bibnamefont
  {Barbo}}, \bibinfo {author} {\bibfnamefont {M.}~\bibnamefont {Bertolo}},
  \bibinfo {author} {\bibfnamefont {A.}~\bibnamefont {Bianco}}, \bibinfo
  {author} {\bibfnamefont {G.}~\bibnamefont {Cautero}}, \bibinfo {author}
  {\bibfnamefont {S.}~\bibnamefont {Fontana}}, \bibinfo {author} {\bibfnamefont
  {T.~K.}\ \bibnamefont {Johal}}, \bibinfo {author} {\bibfnamefont
  {S.}~\bibnamefont {La~Rosa}}, \bibinfo {author} {\bibfnamefont
  {G.}~\bibnamefont {Margaritondo}}, \ and\ \bibinfo {author} {\bibfnamefont
  {K.}~\bibnamefont {Kaznacheyev}},\ }\bibfield  {title} {\enquote {\bibinfo
  {title} {Spectromicroscopy beamline at {ELETTRA}: {Performances} achieved at
  the end of commissioning},}\ }\href {\doibase 10.1063/1.1150148} {\bibfield
  {journal} {\bibinfo  {journal} {Review of Scientific Instruments}\ }\textbf
  {\bibinfo {volume} {71}},\ \bibinfo {pages} {5--10} (\bibinfo {year}
  {2000})}\BibitemShut {NoStop}%
\bibitem [{\citenamefont {Avila}\ \emph {et~al.}(2013)\citenamefont {Avila},
  \citenamefont {Razado-Colambo}, \citenamefont {Lorcy}, \citenamefont
  {Giorgetta}, \citenamefont {Polack},\ and\ \citenamefont
  {Asensio}}]{Avila2013}%
  \BibitemOpen
  \bibfield  {author} {\bibinfo {author} {\bibfnamefont {J.}~\bibnamefont
  {Avila}}, \bibinfo {author} {\bibfnamefont {I.}~\bibnamefont
  {Razado-Colambo}}, \bibinfo {author} {\bibfnamefont {S.}~\bibnamefont
  {Lorcy}}, \bibinfo {author} {\bibfnamefont {J.-L.}\ \bibnamefont
  {Giorgetta}}, \bibinfo {author} {\bibfnamefont {F.}~\bibnamefont {Polack}}, \
  and\ \bibinfo {author} {\bibfnamefont {M.~C.}\ \bibnamefont {Asensio}},\
  }\bibfield  {title} {\enquote {\bibinfo {title} {Interferometer-controlled
  soft {X}-ray scanning photoemission microscope at {SOLEIL}},}\ }\href
  {\doibase 10.1088/1742-6596/425/13/132013} {\bibfield  {journal} {\bibinfo
  {journal} {Journal of Physics: Conference Series}\ }\textbf {\bibinfo
  {volume} {425}},\ \bibinfo {pages} {132013} (\bibinfo {year}
  {2013})}\BibitemShut {NoStop}%
\bibitem [{\citenamefont {Rotenberg}\ and\ \citenamefont
  {Bostwick}(2014)}]{Rotenberg2014}%
  \BibitemOpen
  \bibfield  {author} {\bibinfo {author} {\bibfnamefont {E.}~\bibnamefont
  {Rotenberg}}\ and\ \bibinfo {author} {\bibfnamefont {A.}~\bibnamefont
  {Bostwick}},\ }\bibfield  {title} {\enquote {\bibinfo {title} {{MicroARPES}
  and {nanoARPES} at diffraction-limited light sources: {O}pportunities and
  performance gains},}\ }\href {\doibase 10.1107/S1600577514015409} {\bibfield
  {journal} {\bibinfo  {journal} {Journal of Synchrotron Radiation}\ }\textbf
  {\bibinfo {volume} {21}},\ \bibinfo {pages} {1048--1056} (\bibinfo {year}
  {2014})}\BibitemShut {NoStop}%
\bibitem [{\citenamefont {Faeth}\ \emph {et~al.}(2021)\citenamefont {Faeth},
  \citenamefont {Yang}, \citenamefont {Kawasaki}, \citenamefont {Nelson},
  \citenamefont {Mishra}, \citenamefont {Parzyck}, \citenamefont {Li},
  \citenamefont {Schlom},\ and\ \citenamefont {Shen}}]{Faeth2021}%
  \BibitemOpen
  \bibfield  {author} {\bibinfo {author} {\bibfnamefont {B.}~\bibnamefont
  {Faeth}}, \bibinfo {author} {\bibfnamefont {S.-L.}\ \bibnamefont {Yang}},
  \bibinfo {author} {\bibfnamefont {J.}~\bibnamefont {Kawasaki}}, \bibinfo
  {author} {\bibfnamefont {J.}~\bibnamefont {Nelson}}, \bibinfo {author}
  {\bibfnamefont {P.}~\bibnamefont {Mishra}}, \bibinfo {author} {\bibfnamefont
  {C.}~\bibnamefont {Parzyck}}, \bibinfo {author} {\bibfnamefont
  {C.}~\bibnamefont {Li}}, \bibinfo {author} {\bibfnamefont {D.}~\bibnamefont
  {Schlom}}, \ and\ \bibinfo {author} {\bibfnamefont {K.}~\bibnamefont
  {Shen}},\ }\bibfield  {title} {\enquote {\bibinfo {title} {Incoherent
  {Cooper} pairing and pseudogap behavior in single-layer
  {Fe}{Se}/{Sr}{Ti}{O$_3$}},}\ }\href {\doibase 10.1103/PhysRevX.11.021054}
  {\bibfield  {journal} {\bibinfo  {journal} {Physical Review X}\ }\textbf
  {\bibinfo {volume} {11}},\ \bibinfo {pages} {021054} (\bibinfo {year}
  {2021})}\BibitemShut {NoStop}%
\bibitem [{\citenamefont {Perfetti}\ \emph {et~al.}(2007)\citenamefont
  {Perfetti}, \citenamefont {Loukakos}, \citenamefont {Lisowski}, \citenamefont
  {Bovensiepen}, \citenamefont {Eisaki},\ and\ \citenamefont
  {Wolf}}]{Perfetti2007}%
  \BibitemOpen
  \bibfield  {author} {\bibinfo {author} {\bibfnamefont {L.}~\bibnamefont
  {Perfetti}}, \bibinfo {author} {\bibfnamefont {P.~A.}\ \bibnamefont
  {Loukakos}}, \bibinfo {author} {\bibfnamefont {M.}~\bibnamefont {Lisowski}},
  \bibinfo {author} {\bibfnamefont {U.}~\bibnamefont {Bovensiepen}}, \bibinfo
  {author} {\bibfnamefont {H.}~\bibnamefont {Eisaki}}, \ and\ \bibinfo {author}
  {\bibfnamefont {M.}~\bibnamefont {Wolf}},\ }\bibfield  {title} {\enquote
  {\bibinfo {title} {Ultrafast electron relaxation in superconducting
  {Bi}$_2${Sr}$_2${Ca}{Cu}$_2${O}$_{8+\delta}$ by time-resolved photoelectron
  spectroscopy},}\ }\href {\doibase 10.1103/PhysRevLett.99.197001} {\bibfield
  {journal} {\bibinfo  {journal} {Physical Review Letters}\ }\textbf {\bibinfo
  {volume} {99}},\ \bibinfo {pages} {197001} (\bibinfo {year}
  {2007})}\BibitemShut {NoStop}%
\bibitem [{\citenamefont {Schmitt}\ \emph {et~al.}(2008)\citenamefont
  {Schmitt}, \citenamefont {Kirchmann}, \citenamefont {Bovensiepen},
  \citenamefont {Moore}, \citenamefont {Rettig}, \citenamefont {Krenz},
  \citenamefont {Chu}, \citenamefont {Ru}, \citenamefont {Perfetti},
  \citenamefont {Lu}, \citenamefont {Wolf}, \citenamefont {Fisher},\ and\
  \citenamefont {Shen}}]{Schmitt2008}%
  \BibitemOpen
  \bibfield  {author} {\bibinfo {author} {\bibfnamefont {F.}~\bibnamefont
  {Schmitt}}, \bibinfo {author} {\bibfnamefont {P.~S.}\ \bibnamefont
  {Kirchmann}}, \bibinfo {author} {\bibfnamefont {U.}~\bibnamefont
  {Bovensiepen}}, \bibinfo {author} {\bibfnamefont {R.~G.}\ \bibnamefont
  {Moore}}, \bibinfo {author} {\bibfnamefont {L.}~\bibnamefont {Rettig}},
  \bibinfo {author} {\bibfnamefont {M.}~\bibnamefont {Krenz}}, \bibinfo
  {author} {\bibfnamefont {J.-H.}\ \bibnamefont {Chu}}, \bibinfo {author}
  {\bibfnamefont {N.}~\bibnamefont {Ru}}, \bibinfo {author} {\bibfnamefont
  {L.}~\bibnamefont {Perfetti}}, \bibinfo {author} {\bibfnamefont {D.~H.}\
  \bibnamefont {Lu}}, \bibinfo {author} {\bibfnamefont {M.}~\bibnamefont
  {Wolf}}, \bibinfo {author} {\bibfnamefont {I.~R.}\ \bibnamefont {Fisher}}, \
  and\ \bibinfo {author} {\bibfnamefont {Z.-X.}\ \bibnamefont {Shen}},\
  }\bibfield  {title} {\enquote {\bibinfo {title} {Transient electronic
  structure and melting of a charge density wave in {Tb}{Te$_3$}},}\ }\href
  {https://www.science.org/doi/abs/10.1126/science.1160778} {\bibfield
  {journal} {\bibinfo  {journal} {Science}\ } (\bibinfo {year}
  {2008})}\BibitemShut {NoStop}%
\bibitem [{\citenamefont {Sobota}\ \emph {et~al.}(2012)\citenamefont {Sobota},
  \citenamefont {Yang}, \citenamefont {Analytis}, \citenamefont {Chen},
  \citenamefont {Fisher}, \citenamefont {Kirchmann},\ and\ \citenamefont
  {Shen}}]{Sobota2012}%
  \BibitemOpen
  \bibfield  {author} {\bibinfo {author} {\bibfnamefont {J.~A.}\ \bibnamefont
  {Sobota}}, \bibinfo {author} {\bibfnamefont {S.}~\bibnamefont {Yang}},
  \bibinfo {author} {\bibfnamefont {J.~G.}\ \bibnamefont {Analytis}}, \bibinfo
  {author} {\bibfnamefont {Y.~L.}\ \bibnamefont {Chen}}, \bibinfo {author}
  {\bibfnamefont {I.~R.}\ \bibnamefont {Fisher}}, \bibinfo {author}
  {\bibfnamefont {P.~S.}\ \bibnamefont {Kirchmann}}, \ and\ \bibinfo {author}
  {\bibfnamefont {Z.-X.}\ \bibnamefont {Shen}},\ }\bibfield  {title} {\enquote
  {\bibinfo {title} {Ultrafast optical excitation of a persistent surface-state
  population in the topological insulator {Bi$_2$}{Se$_3$}},}\ }\href {\doibase
  10.1103/PhysRevLett.108.117403} {\bibfield  {journal} {\bibinfo  {journal}
  {Physical Review Letters}\ }\textbf {\bibinfo {volume} {108}},\ \bibinfo
  {pages} {117403} (\bibinfo {year} {2012})}\BibitemShut {NoStop}%
\bibitem [{\citenamefont {Smallwood}\ \emph {et~al.}(2012)\citenamefont
  {Smallwood}, \citenamefont {Hinton}, \citenamefont {Jozwiak}, \citenamefont
  {Zhang}, \citenamefont {Koralek}, \citenamefont {Eisaki}, \citenamefont
  {Lee}, \citenamefont {Orenstein},\ and\ \citenamefont
  {Lanzara}}]{Smallwood2012}%
  \BibitemOpen
  \bibfield  {author} {\bibinfo {author} {\bibfnamefont {C.~L.}\ \bibnamefont
  {Smallwood}}, \bibinfo {author} {\bibfnamefont {J.~P.}\ \bibnamefont
  {Hinton}}, \bibinfo {author} {\bibfnamefont {C.}~\bibnamefont {Jozwiak}},
  \bibinfo {author} {\bibfnamefont {W.}~\bibnamefont {Zhang}}, \bibinfo
  {author} {\bibfnamefont {J.~D.}\ \bibnamefont {Koralek}}, \bibinfo {author}
  {\bibfnamefont {H.}~\bibnamefont {Eisaki}}, \bibinfo {author} {\bibfnamefont
  {D.-H.}\ \bibnamefont {Lee}}, \bibinfo {author} {\bibfnamefont
  {J.}~\bibnamefont {Orenstein}}, \ and\ \bibinfo {author} {\bibfnamefont
  {A.}~\bibnamefont {Lanzara}},\ }\bibfield  {title} {\enquote {\bibinfo
  {title} {Tracking {Cooper} pairs in a cuprate superconductor by ultrafast
  angle-resolved photoemission},}\ }\href
  {https://www.science.org/doi/abs/10.1126/science.1217423} {\bibfield
  {journal} {\bibinfo  {journal} {Science}\ } (\bibinfo {year}
  {2012})}\BibitemShut {NoStop}%
\bibitem [{\citenamefont {Wang}\ \emph {et~al.}(2013)\citenamefont {Wang},
  \citenamefont {Steinberg}, \citenamefont {Jarillo-Herrero},\ and\
  \citenamefont {Gedik}}]{Wang2013}%
  \BibitemOpen
  \bibfield  {author} {\bibinfo {author} {\bibfnamefont {Y.~H.}\ \bibnamefont
  {Wang}}, \bibinfo {author} {\bibfnamefont {H.}~\bibnamefont {Steinberg}},
  \bibinfo {author} {\bibfnamefont {P.}~\bibnamefont {Jarillo-Herrero}}, \ and\
  \bibinfo {author} {\bibfnamefont {N.}~\bibnamefont {Gedik}},\ }\bibfield
  {title} {\enquote {\bibinfo {title} {Observation of {Floquet}-{Bloch} states
  on the surface of a topological insulator},}\ }\href
  {https://www.science.org/doi/abs/10.1126/science.1239834} {\bibfield
  {journal} {\bibinfo  {journal} {Science}\ } (\bibinfo {year}
  {2013})}\BibitemShut {NoStop}%
\bibitem [{\citenamefont {Yang}\ \emph
  {et~al.}(2015{\natexlab{a}})\citenamefont {Yang}, \citenamefont {Sobota},
  \citenamefont {Leuenberger}, \citenamefont {He}, \citenamefont {Hashimoto},
  \citenamefont {Lu}, \citenamefont {Eisaki}, \citenamefont {Kirchmann},\ and\
  \citenamefont {Shen}}]{Yang2015inequivalence}%
  \BibitemOpen
  \bibfield  {author} {\bibinfo {author} {\bibfnamefont {S.-L.}\ \bibnamefont
  {Yang}}, \bibinfo {author} {\bibfnamefont {J.}~\bibnamefont {Sobota}},
  \bibinfo {author} {\bibfnamefont {D.}~\bibnamefont {Leuenberger}}, \bibinfo
  {author} {\bibfnamefont {Y.}~\bibnamefont {He}}, \bibinfo {author}
  {\bibfnamefont {M.}~\bibnamefont {Hashimoto}}, \bibinfo {author}
  {\bibfnamefont {D.}~\bibnamefont {Lu}}, \bibinfo {author} {\bibfnamefont
  {H.}~\bibnamefont {Eisaki}}, \bibinfo {author} {\bibfnamefont
  {P.}~\bibnamefont {Kirchmann}}, \ and\ \bibinfo {author} {\bibfnamefont
  {Z.-X.}\ \bibnamefont {Shen}},\ }\bibfield  {title} {\enquote {\bibinfo
  {title} {Inequivalence of single-particle and population lifetimes in a
  cuprate superconductor},}\ }\href {\doibase 10.1103/PhysRevLett.114.247001}
  {\bibfield  {journal} {\bibinfo  {journal} {Physical Review Letters}\
  }\textbf {\bibinfo {volume} {114}},\ \bibinfo {pages} {247001} (\bibinfo
  {year} {2015}{\natexlab{a}})}\BibitemShut {NoStop}%
\bibitem [{\citenamefont {Gerber}\ \emph {et~al.}(2017)\citenamefont {Gerber},
  \citenamefont {Yang}, \citenamefont {Zhu}, \citenamefont {Soifer},
  \citenamefont {Sobota}, \citenamefont {Rebec}, \citenamefont {Lee},
  \citenamefont {Jia}, \citenamefont {Moritz}, \citenamefont {Jia},
  \citenamefont {Gauthier}, \citenamefont {Li}, \citenamefont {Leuenberger},
  \citenamefont {Zhang}, \citenamefont {Chaix}, \citenamefont {Li},
  \citenamefont {Jang}, \citenamefont {Lee}, \citenamefont {Yi}, \citenamefont
  {Dakovski}, \citenamefont {Song}, \citenamefont {Glownia}, \citenamefont
  {Nelson}, \citenamefont {Kim}, \citenamefont {Chuang}, \citenamefont
  {Hussain}, \citenamefont {Moore}, \citenamefont {Devereaux}, \citenamefont
  {Lee}, \citenamefont {Kirchmann},\ and\ \citenamefont {Shen}}]{Gerber2017}%
  \BibitemOpen
  \bibfield  {author} {\bibinfo {author} {\bibfnamefont {S.}~\bibnamefont
  {Gerber}}, \bibinfo {author} {\bibfnamefont {S.-L.}\ \bibnamefont {Yang}},
  \bibinfo {author} {\bibfnamefont {D.}~\bibnamefont {Zhu}}, \bibinfo {author}
  {\bibfnamefont {H.}~\bibnamefont {Soifer}}, \bibinfo {author} {\bibfnamefont
  {J.~A.}\ \bibnamefont {Sobota}}, \bibinfo {author} {\bibfnamefont
  {S.}~\bibnamefont {Rebec}}, \bibinfo {author} {\bibfnamefont {J.~J.}\
  \bibnamefont {Lee}}, \bibinfo {author} {\bibfnamefont {T.}~\bibnamefont
  {Jia}}, \bibinfo {author} {\bibfnamefont {B.}~\bibnamefont {Moritz}},
  \bibinfo {author} {\bibfnamefont {C.}~\bibnamefont {Jia}}, \bibinfo {author}
  {\bibfnamefont {A.}~\bibnamefont {Gauthier}}, \bibinfo {author}
  {\bibfnamefont {Y.}~\bibnamefont {Li}}, \bibinfo {author} {\bibfnamefont
  {D.}~\bibnamefont {Leuenberger}}, \bibinfo {author} {\bibfnamefont
  {Y.}~\bibnamefont {Zhang}}, \bibinfo {author} {\bibfnamefont
  {L.}~\bibnamefont {Chaix}}, \bibinfo {author} {\bibfnamefont
  {W.}~\bibnamefont {Li}}, \bibinfo {author} {\bibfnamefont {H.}~\bibnamefont
  {Jang}}, \bibinfo {author} {\bibfnamefont {J.-S.}\ \bibnamefont {Lee}},
  \bibinfo {author} {\bibfnamefont {M.}~\bibnamefont {Yi}}, \bibinfo {author}
  {\bibfnamefont {G.~L.}\ \bibnamefont {Dakovski}}, \bibinfo {author}
  {\bibfnamefont {S.}~\bibnamefont {Song}}, \bibinfo {author} {\bibfnamefont
  {J.~M.}\ \bibnamefont {Glownia}}, \bibinfo {author} {\bibfnamefont
  {S.}~\bibnamefont {Nelson}}, \bibinfo {author} {\bibfnamefont {K.~W.}\
  \bibnamefont {Kim}}, \bibinfo {author} {\bibfnamefont {Y.-D.}\ \bibnamefont
  {Chuang}}, \bibinfo {author} {\bibfnamefont {Z.}~\bibnamefont {Hussain}},
  \bibinfo {author} {\bibfnamefont {R.~G.}\ \bibnamefont {Moore}}, \bibinfo
  {author} {\bibfnamefont {T.~P.}\ \bibnamefont {Devereaux}}, \bibinfo {author}
  {\bibfnamefont {W.-S.}\ \bibnamefont {Lee}}, \bibinfo {author} {\bibfnamefont
  {P.~S.}\ \bibnamefont {Kirchmann}}, \ and\ \bibinfo {author} {\bibfnamefont
  {Z.-X.}\ \bibnamefont {Shen}},\ }\bibfield  {title} {\enquote {\bibinfo
  {title} {Femtosecond electron-phonon lock-in by photoemission and {X}-ray
  free-electron laser},}\ }\href {\doibase 10.1126/science.aak9946} {\bibfield
  {journal} {\bibinfo  {journal} {Science}\ }\textbf {\bibinfo {volume}
  {357}},\ \bibinfo {pages} {71--75} (\bibinfo {year} {2017})}\BibitemShut
  {NoStop}%
\bibitem [{\citenamefont {Sobota}\ \emph {et~al.}(2013)\citenamefont {Sobota},
  \citenamefont {Yang}, \citenamefont {Kemper}, \citenamefont {Lee},
  \citenamefont {Schmitt}, \citenamefont {Li}, \citenamefont {Moore},
  \citenamefont {Analytis}, \citenamefont {Fisher}, \citenamefont {Kirchmann},
  \citenamefont {Devereaux},\ and\ \citenamefont {Shen}}]{Sobota2013}%
  \BibitemOpen
  \bibfield  {author} {\bibinfo {author} {\bibfnamefont {J.~A.}\ \bibnamefont
  {Sobota}}, \bibinfo {author} {\bibfnamefont {S.-L.}\ \bibnamefont {Yang}},
  \bibinfo {author} {\bibfnamefont {A.~F.}\ \bibnamefont {Kemper}}, \bibinfo
  {author} {\bibfnamefont {J.~J.}\ \bibnamefont {Lee}}, \bibinfo {author}
  {\bibfnamefont {F.~T.}\ \bibnamefont {Schmitt}}, \bibinfo {author}
  {\bibfnamefont {W.}~\bibnamefont {Li}}, \bibinfo {author} {\bibfnamefont
  {R.~G.}\ \bibnamefont {Moore}}, \bibinfo {author} {\bibfnamefont {J.~G.}\
  \bibnamefont {Analytis}}, \bibinfo {author} {\bibfnamefont {I.~R.}\
  \bibnamefont {Fisher}}, \bibinfo {author} {\bibfnamefont {P.~S.}\
  \bibnamefont {Kirchmann}}, \bibinfo {author} {\bibfnamefont {T.~P.}\
  \bibnamefont {Devereaux}}, \ and\ \bibinfo {author} {\bibfnamefont {Z.-X.}\
  \bibnamefont {Shen}},\ }\bibfield  {title} {\enquote {\bibinfo {title}
  {Direct optical coupling to an unoccupied {D}irac surface state in the
  topological insulator {Bi$_2$Se$_3$}},}\ }\href {\doibase
  10.1103/PhysRevLett.111.136802} {\bibfield  {journal} {\bibinfo  {journal}
  {Phys. Rev. Lett.}\ }\textbf {\bibinfo {volume} {111}},\ \bibinfo {pages}
  {136802} (\bibinfo {year} {2013})}\BibitemShut {NoStop}%
\bibitem [{\citenamefont {Soifer}\ \emph {et~al.}(2019)\citenamefont {Soifer},
  \citenamefont {Gauthier}, \citenamefont {Kemper}, \citenamefont {Rotundu},
  \citenamefont {Yang}, \citenamefont {Xiong}, \citenamefont {Lu},
  \citenamefont {Hashimoto}, \citenamefont {Kirchmann}, \citenamefont
  {Sobota},\ and\ \citenamefont {Shen}}]{Soifer2019}%
  \BibitemOpen
  \bibfield  {author} {\bibinfo {author} {\bibfnamefont {H.}~\bibnamefont
  {Soifer}}, \bibinfo {author} {\bibfnamefont {A.}~\bibnamefont {Gauthier}},
  \bibinfo {author} {\bibfnamefont {A.}~\bibnamefont {Kemper}}, \bibinfo
  {author} {\bibfnamefont {C.}~\bibnamefont {Rotundu}}, \bibinfo {author}
  {\bibfnamefont {S.-L.}\ \bibnamefont {Yang}}, \bibinfo {author}
  {\bibfnamefont {H.}~\bibnamefont {Xiong}}, \bibinfo {author} {\bibfnamefont
  {D.}~\bibnamefont {Lu}}, \bibinfo {author} {\bibfnamefont {M.}~\bibnamefont
  {Hashimoto}}, \bibinfo {author} {\bibfnamefont {P.}~\bibnamefont
  {Kirchmann}}, \bibinfo {author} {\bibfnamefont {J.}~\bibnamefont {Sobota}}, \
  and\ \bibinfo {author} {\bibfnamefont {Z.-X.}\ \bibnamefont {Shen}},\
  }\bibfield  {title} {\enquote {\bibinfo {title} {Band-resolved imaging of
  photocurrent in a topological insulator},}\ }\href {\doibase
  10.1103/PhysRevLett.122.167401} {\bibfield  {journal} {\bibinfo  {journal}
  {Physical Review Letters}\ }\textbf {\bibinfo {volume} {122}},\ \bibinfo
  {pages} {167401} (\bibinfo {year} {2019})}\BibitemShut {NoStop}%
\bibitem [{\citenamefont {Gauthier}\ \emph {et~al.}(2020)\citenamefont
  {Gauthier}, \citenamefont {Sobota}, \citenamefont {Gauthier}, \citenamefont
  {Xu}, \citenamefont {Pfau}, \citenamefont {Rotundu}, \citenamefont {Shen},\
  and\ \citenamefont {Kirchmann}}]{Gauthier2020}%
  \BibitemOpen
  \bibfield  {author} {\bibinfo {author} {\bibfnamefont {A.}~\bibnamefont
  {Gauthier}}, \bibinfo {author} {\bibfnamefont {J.~A.}\ \bibnamefont
  {Sobota}}, \bibinfo {author} {\bibfnamefont {N.}~\bibnamefont {Gauthier}},
  \bibinfo {author} {\bibfnamefont {K.-J.}\ \bibnamefont {Xu}}, \bibinfo
  {author} {\bibfnamefont {H.}~\bibnamefont {Pfau}}, \bibinfo {author}
  {\bibfnamefont {C.~R.}\ \bibnamefont {Rotundu}}, \bibinfo {author}
  {\bibfnamefont {Z.-X.}\ \bibnamefont {Shen}}, \ and\ \bibinfo {author}
  {\bibfnamefont {P.~S.}\ \bibnamefont {Kirchmann}},\ }\bibfield  {title}
  {\enquote {\bibinfo {title} {Tuning time and energy resolution in
  time-resolved photoemission spectroscopy with nonlinear crystals},}\ }\href
  {\doibase 10.1063/5.0018834} {\bibfield  {journal} {\bibinfo  {journal}
  {Journal of Applied Physics}\ }\textbf {\bibinfo {volume} {128}},\ \bibinfo
  {pages} {093101} (\bibinfo {year} {2020})}\BibitemShut {NoStop}%
\bibitem [{\citenamefont {He}\ \emph {et~al.}(2016)\citenamefont {He},
  \citenamefont {Vishik}, \citenamefont {Yi}, \citenamefont {Yang},
  \citenamefont {Liu}, \citenamefont {Lee}, \citenamefont {Chen}, \citenamefont
  {Rebec}, \citenamefont {Leuenberger}, \citenamefont {Zong}, \citenamefont
  {Jefferson}, \citenamefont {Moore}, \citenamefont {Kirchmann}, \citenamefont
  {Merriam},\ and\ \citenamefont {Shen}}]{He2016}%
  \BibitemOpen
  \bibfield  {author} {\bibinfo {author} {\bibfnamefont {Y.}~\bibnamefont
  {He}}, \bibinfo {author} {\bibfnamefont {I.~M.}\ \bibnamefont {Vishik}},
  \bibinfo {author} {\bibfnamefont {M.}~\bibnamefont {Yi}}, \bibinfo {author}
  {\bibfnamefont {S.}~\bibnamefont {Yang}}, \bibinfo {author} {\bibfnamefont
  {Z.}~\bibnamefont {Liu}}, \bibinfo {author} {\bibfnamefont {J.~J.}\
  \bibnamefont {Lee}}, \bibinfo {author} {\bibfnamefont {S.}~\bibnamefont
  {Chen}}, \bibinfo {author} {\bibfnamefont {S.~N.}\ \bibnamefont {Rebec}},
  \bibinfo {author} {\bibfnamefont {D.}~\bibnamefont {Leuenberger}}, \bibinfo
  {author} {\bibfnamefont {A.}~\bibnamefont {Zong}}, \bibinfo {author}
  {\bibfnamefont {C.~M.}\ \bibnamefont {Jefferson}}, \bibinfo {author}
  {\bibfnamefont {R.~G.}\ \bibnamefont {Moore}}, \bibinfo {author}
  {\bibfnamefont {P.~S.}\ \bibnamefont {Kirchmann}}, \bibinfo {author}
  {\bibfnamefont {A.~J.}\ \bibnamefont {Merriam}}, \ and\ \bibinfo {author}
  {\bibfnamefont {Z.-X.}\ \bibnamefont {Shen}},\ }\bibfield  {title} {\enquote
  {\bibinfo {title} {Invited {Article}: {High} resolution angle resolved
  photoemission with tabletop 11 {eV} laser},}\ }\href {\doibase
  10.1063/1.4939759} {\bibfield  {journal} {\bibinfo  {journal} {Review of
  Scientific Instruments}\ }\textbf {\bibinfo {volume} {87}},\ \bibinfo {pages}
  {011301} (\bibinfo {year} {2016})}\BibitemShut {NoStop}%
\bibitem [{\citenamefont {Corder}\ \emph {et~al.}(2018)\citenamefont {Corder},
  \citenamefont {Zhao}, \citenamefont {Bakalis}, \citenamefont {Li},
  \citenamefont {Kershis}, \citenamefont {Muraca}, \citenamefont {White},\ and\
  \citenamefont {Allison}}]{Corder2018}%
  \BibitemOpen
  \bibfield  {author} {\bibinfo {author} {\bibfnamefont {C.}~\bibnamefont
  {Corder}}, \bibinfo {author} {\bibfnamefont {P.}~\bibnamefont {Zhao}},
  \bibinfo {author} {\bibfnamefont {J.}~\bibnamefont {Bakalis}}, \bibinfo
  {author} {\bibfnamefont {X.}~\bibnamefont {Li}}, \bibinfo {author}
  {\bibfnamefont {M.~D.}\ \bibnamefont {Kershis}}, \bibinfo {author}
  {\bibfnamefont {A.~R.}\ \bibnamefont {Muraca}}, \bibinfo {author}
  {\bibfnamefont {M.~G.}\ \bibnamefont {White}}, \ and\ \bibinfo {author}
  {\bibfnamefont {T.~K.}\ \bibnamefont {Allison}},\ }\bibfield  {title}
  {\enquote {\bibinfo {title} {Ultrafast extreme ultraviolet photoemission
  without space charge},}\ }\href {\doibase 10.1063/1.5045578} {\bibfield
  {journal} {\bibinfo  {journal} {Structural Dynamics}\ }\textbf {\bibinfo
  {volume} {5}},\ \bibinfo {pages} {054301} (\bibinfo {year}
  {2018})}\BibitemShut {NoStop}%
\bibitem [{\citenamefont {Zhou}\ \emph {et~al.}(2003)\citenamefont {Zhou},
  \citenamefont {Johnson}, \citenamefont {Hone},\ and\ \citenamefont
  {Smith}}]{Zhou2003}%
  \BibitemOpen
  \bibfield  {author} {\bibinfo {author} {\bibfnamefont {Y.~X.}\ \bibnamefont
  {Zhou}}, \bibinfo {author} {\bibfnamefont {A.~T.}\ \bibnamefont {Johnson}},
  \bibinfo {author} {\bibfnamefont {J.}~\bibnamefont {Hone}}, \ and\ \bibinfo
  {author} {\bibfnamefont {W.~F.}\ \bibnamefont {Smith}},\ }\bibfield  {title}
  {\enquote {\bibinfo {title} {Simple fabrication of molecular circuits by
  shadow mask evaporation},}\ }\href {\doibase 10.1021/nl034512y} {\bibfield
  {journal} {\bibinfo  {journal} {Nano Letters}\ }\textbf {\bibinfo {volume}
  {3}},\ \bibinfo {pages} {1371--1374} (\bibinfo {year} {2003})}\BibitemShut
  {NoStop}%
\bibitem [{\citenamefont {Tsioutsios}\ \emph {et~al.}(2020)\citenamefont
  {Tsioutsios}, \citenamefont {Serniak}, \citenamefont {Diamond}, \citenamefont
  {Sivak}, \citenamefont {Wang}, \citenamefont {Shankar}, \citenamefont
  {Frunzio}, \citenamefont {Schoelkopf},\ and\ \citenamefont
  {Devoret}}]{Tsioutsios2020}%
  \BibitemOpen
  \bibfield  {author} {\bibinfo {author} {\bibfnamefont {I.}~\bibnamefont
  {Tsioutsios}}, \bibinfo {author} {\bibfnamefont {K.}~\bibnamefont {Serniak}},
  \bibinfo {author} {\bibfnamefont {S.}~\bibnamefont {Diamond}}, \bibinfo
  {author} {\bibfnamefont {V.~V.}\ \bibnamefont {Sivak}}, \bibinfo {author}
  {\bibfnamefont {Z.}~\bibnamefont {Wang}}, \bibinfo {author} {\bibfnamefont
  {S.}~\bibnamefont {Shankar}}, \bibinfo {author} {\bibfnamefont
  {L.}~\bibnamefont {Frunzio}}, \bibinfo {author} {\bibfnamefont {R.~J.}\
  \bibnamefont {Schoelkopf}}, \ and\ \bibinfo {author} {\bibfnamefont {M.~H.}\
  \bibnamefont {Devoret}},\ }\bibfield  {title} {\enquote {\bibinfo {title}
  {Free-standing silicon shadow masks for transmon qubit fabrication},}\ }\href
  {\doibase 10.1063/1.5138953} {\bibfield  {journal} {\bibinfo  {journal} {AIP
  Advances}\ }\textbf {\bibinfo {volume} {10}},\ \bibinfo {pages} {065120}
  (\bibinfo {year} {2020})}\BibitemShut {NoStop}%
\bibitem [{\citenamefont {Wolfgang}(2014)}]{Demtroder2014}%
  \BibitemOpen
  \bibfield  {author} {\bibinfo {author} {\bibfnamefont {D.}~\bibnamefont
  {Wolfgang}},\ }\href@noop {} {\emph {\bibinfo {title} {Laser {S}pectroscopy
  1: {B}asic {P}rinciples}}}\ (\bibinfo  {publisher} {Springer},\ \bibinfo
  {year} {2014})\BibitemShut {NoStop}%
\bibitem [{\citenamefont {Zhou}\ \emph {et~al.}(2005)\citenamefont {Zhou},
  \citenamefont {Wannberg}, \citenamefont {Yang}, \citenamefont {Brouet},
  \citenamefont {Sun}, \citenamefont {Douglas}, \citenamefont {Dessau},
  \citenamefont {Hussain},\ and\ \citenamefont {Shen}}]{Zhou2005}%
  \BibitemOpen
  \bibfield  {author} {\bibinfo {author} {\bibfnamefont {X.}~\bibnamefont
  {Zhou}}, \bibinfo {author} {\bibfnamefont {B.}~\bibnamefont {Wannberg}},
  \bibinfo {author} {\bibfnamefont {W.}~\bibnamefont {Yang}}, \bibinfo {author}
  {\bibfnamefont {V.}~\bibnamefont {Brouet}}, \bibinfo {author} {\bibfnamefont
  {Z.}~\bibnamefont {Sun}}, \bibinfo {author} {\bibfnamefont {J.}~\bibnamefont
  {Douglas}}, \bibinfo {author} {\bibfnamefont {D.}~\bibnamefont {Dessau}},
  \bibinfo {author} {\bibfnamefont {Z.}~\bibnamefont {Hussain}}, \ and\
  \bibinfo {author} {\bibfnamefont {Z.-X.}\ \bibnamefont {Shen}},\ }\bibfield
  {title} {\enquote {\bibinfo {title} {Space charge effect and mirror charge
  effect in photoemission spectroscopy},}\ }\href {\doibase
  https://doi.org/10.1016/j.elspec.2004.08.004} {\bibfield  {journal} {\bibinfo
   {journal} {Journal of Electron Spectroscopy and Related Phenomena}\ }\textbf
  {\bibinfo {volume} {142}},\ \bibinfo {pages} {27--38} (\bibinfo {year}
  {2005})}\BibitemShut {NoStop}%
\bibitem [{\citenamefont {Ishida}\ \emph {et~al.}(2014)\citenamefont {Ishida},
  \citenamefont {Togashi}, \citenamefont {Yamamoto}, \citenamefont {Tanaka},
  \citenamefont {Kiss}, \citenamefont {Otsu}, \citenamefont {Kobayashi},\ and\
  \citenamefont {Shin}}]{Ishida2014}%
  \BibitemOpen
  \bibfield  {author} {\bibinfo {author} {\bibfnamefont {Y.}~\bibnamefont
  {Ishida}}, \bibinfo {author} {\bibfnamefont {T.}~\bibnamefont {Togashi}},
  \bibinfo {author} {\bibfnamefont {K.}~\bibnamefont {Yamamoto}}, \bibinfo
  {author} {\bibfnamefont {M.}~\bibnamefont {Tanaka}}, \bibinfo {author}
  {\bibfnamefont {T.}~\bibnamefont {Kiss}}, \bibinfo {author} {\bibfnamefont
  {T.}~\bibnamefont {Otsu}}, \bibinfo {author} {\bibfnamefont {Y.}~\bibnamefont
  {Kobayashi}}, \ and\ \bibinfo {author} {\bibfnamefont {S.}~\bibnamefont
  {Shin}},\ }\bibfield  {title} {\enquote {\bibinfo {title} {Time-resolved
  photoemission apparatus achieving sub-20-{meV} energy resolution and high
  stability},}\ }\href {\doibase 10.1063/1.4903788} {\bibfield  {journal}
  {\bibinfo  {journal} {Review of Scientific Instruments}\ }\textbf {\bibinfo
  {volume} {85}},\ \bibinfo {pages} {123904} (\bibinfo {year}
  {2014})}\BibitemShut {NoStop}%
\bibitem [{\citenamefont {Yang}\ \emph {et~al.}(2019)\citenamefont {Yang},
  \citenamefont {Tang}, \citenamefont {Duan}, \citenamefont {Zhou},
  \citenamefont {Hao},\ and\ \citenamefont {Zhang}}]{Yang2019}%
  \BibitemOpen
  \bibfield  {author} {\bibinfo {author} {\bibfnamefont {Y.}~\bibnamefont
  {Yang}}, \bibinfo {author} {\bibfnamefont {T.}~\bibnamefont {Tang}}, \bibinfo
  {author} {\bibfnamefont {S.}~\bibnamefont {Duan}}, \bibinfo {author}
  {\bibfnamefont {C.}~\bibnamefont {Zhou}}, \bibinfo {author} {\bibfnamefont
  {D.}~\bibnamefont {Hao}}, \ and\ \bibinfo {author} {\bibfnamefont
  {W.}~\bibnamefont {Zhang}},\ }\bibfield  {title} {\enquote {\bibinfo {title}
  {A time- and angle-resolved photoemission spectroscopy with probe photon
  energy up to 6.7 {eV}},}\ }\href {\doibase 10.1063/1.5090439} {\bibfield
  {journal} {\bibinfo  {journal} {Review of Scientific Instruments}\ }\textbf
  {\bibinfo {volume} {90}},\ \bibinfo {pages} {063905} (\bibinfo {year}
  {2019})}\BibitemShut {NoStop}%
\bibitem [{\citenamefont {Graf}\ \emph {et~al.}(2011)\citenamefont {Graf},
  \citenamefont {Jozwiak}, \citenamefont {Smallwood}, \citenamefont {Eisaki},
  \citenamefont {Kaindl}, \citenamefont {Lee},\ and\ \citenamefont
  {Lanzara}}]{Graf2011}%
  \BibitemOpen
  \bibfield  {author} {\bibinfo {author} {\bibfnamefont {J.}~\bibnamefont
  {Graf}}, \bibinfo {author} {\bibfnamefont {C.}~\bibnamefont {Jozwiak}},
  \bibinfo {author} {\bibfnamefont {C.~L.}\ \bibnamefont {Smallwood}}, \bibinfo
  {author} {\bibfnamefont {H.}~\bibnamefont {Eisaki}}, \bibinfo {author}
  {\bibfnamefont {R.~A.}\ \bibnamefont {Kaindl}}, \bibinfo {author}
  {\bibfnamefont {D.-H.}\ \bibnamefont {Lee}}, \ and\ \bibinfo {author}
  {\bibfnamefont {A.}~\bibnamefont {Lanzara}},\ }\bibfield  {title} {\enquote
  {\bibinfo {title} {Nodal quasiparticle meltdown in ultrahigh-resolution
  pump–probe angle-resolved photoemission},}\ }\href@noop {} {\bibfield
  {journal} {\bibinfo  {journal} {Nature Physics}\ }\textbf {\bibinfo {volume}
  {7}} (\bibinfo {year} {2011})}\BibitemShut {NoStop}%
\bibitem [{\citenamefont {Wang}\ \emph
  {et~al.}(2012{\natexlab{a}})\citenamefont {Wang}, \citenamefont {Hsieh},
  \citenamefont {Sie}, \citenamefont {Steinberg}, \citenamefont {Gardner},
  \citenamefont {Lee}, \citenamefont {Jarillo-Herrero},\ and\ \citenamefont
  {Gedik}}]{Wang2012a}%
  \BibitemOpen
  \bibfield  {author} {\bibinfo {author} {\bibfnamefont {Y.~H.}\ \bibnamefont
  {Wang}}, \bibinfo {author} {\bibfnamefont {D.}~\bibnamefont {Hsieh}},
  \bibinfo {author} {\bibfnamefont {E.~J.}\ \bibnamefont {Sie}}, \bibinfo
  {author} {\bibfnamefont {H.}~\bibnamefont {Steinberg}}, \bibinfo {author}
  {\bibfnamefont {D.~R.}\ \bibnamefont {Gardner}}, \bibinfo {author}
  {\bibfnamefont {Y.~S.}\ \bibnamefont {Lee}}, \bibinfo {author} {\bibfnamefont
  {P.}~\bibnamefont {Jarillo-Herrero}}, \ and\ \bibinfo {author} {\bibfnamefont
  {N.}~\bibnamefont {Gedik}},\ }\bibfield  {title} {\enquote {\bibinfo {title}
  {Measurement of intrinsic {Dirac} fermion cooling on the surface of the
  topological insulator {Bi$_2$Se$_3$} using time-resolved and angle-resolved
  photoemission spectroscopy},}\ }\href {\doibase
  10.1103/PhysRevLett.109.127401} {\bibfield  {journal} {\bibinfo  {journal}
  {Physical Review Letters}\ }\textbf {\bibinfo {volume} {109}},\ \bibinfo
  {pages} {127401} (\bibinfo {year} {2012}{\natexlab{a}})}\BibitemShut
  {NoStop}%
\bibitem [{\citenamefont {Wang}\ \emph
  {et~al.}(2012{\natexlab{b}})\citenamefont {Wang}, \citenamefont {Li},
  \citenamefont {Zhang}, \citenamefont {Zhang}, \citenamefont {Zhang},
  \citenamefont {Li}, \citenamefont {Ding}, \citenamefont {Ou}, \citenamefont
  {Deng}, \citenamefont {Chang}, \citenamefont {Wen}, \citenamefont {Song},
  \citenamefont {He}, \citenamefont {Jia}, \citenamefont {Ji}, \citenamefont
  {Wang}, \citenamefont {Wang}, \citenamefont {Chen}, \citenamefont {Ma},\ and\
  \citenamefont {Xue}}]{Wang2012}%
  \BibitemOpen
  \bibfield  {author} {\bibinfo {author} {\bibfnamefont {Q.-Y.}\ \bibnamefont
  {Wang}}, \bibinfo {author} {\bibfnamefont {Z.}~\bibnamefont {Li}}, \bibinfo
  {author} {\bibfnamefont {W.-H.}\ \bibnamefont {Zhang}}, \bibinfo {author}
  {\bibfnamefont {Z.-C.}\ \bibnamefont {Zhang}}, \bibinfo {author}
  {\bibfnamefont {J.-S.}\ \bibnamefont {Zhang}}, \bibinfo {author}
  {\bibfnamefont {W.}~\bibnamefont {Li}}, \bibinfo {author} {\bibfnamefont
  {H.}~\bibnamefont {Ding}}, \bibinfo {author} {\bibfnamefont {Y.-B.}\
  \bibnamefont {Ou}}, \bibinfo {author} {\bibfnamefont {P.}~\bibnamefont
  {Deng}}, \bibinfo {author} {\bibfnamefont {K.}~\bibnamefont {Chang}},
  \bibinfo {author} {\bibfnamefont {J.}~\bibnamefont {Wen}}, \bibinfo {author}
  {\bibfnamefont {C.-L.}\ \bibnamefont {Song}}, \bibinfo {author}
  {\bibfnamefont {K.}~\bibnamefont {He}}, \bibinfo {author} {\bibfnamefont
  {J.-F.}\ \bibnamefont {Jia}}, \bibinfo {author} {\bibfnamefont {S.-H.}\
  \bibnamefont {Ji}}, \bibinfo {author} {\bibfnamefont {Y.-Y.}\ \bibnamefont
  {Wang}}, \bibinfo {author} {\bibfnamefont {L.-L.}\ \bibnamefont {Wang}},
  \bibinfo {author} {\bibfnamefont {X.}~\bibnamefont {Chen}}, \bibinfo {author}
  {\bibfnamefont {X.-C.}\ \bibnamefont {Ma}}, \ and\ \bibinfo {author}
  {\bibfnamefont {Q.-K.}\ \bibnamefont {Xue}},\ }\bibfield  {title} {\enquote
  {\bibinfo {title} {Interface-induced high-temperature superconductivity in
  single unit-cell {FeSe} films on {SrTiO}$_3$},}\ }\href {\doibase
  10.1088/0256-307x/29/3/037402} {\bibfield  {journal} {\bibinfo  {journal}
  {Chinese Physics Letters}\ }\textbf {\bibinfo {volume} {29}},\ \bibinfo
  {pages} {037402} (\bibinfo {year} {2012}{\natexlab{b}})}\BibitemShut
  {NoStop}%
\bibitem [{\citenamefont {Lee}\ \emph {et~al.}(2014)\citenamefont {Lee},
  \citenamefont {Schmitt}, \citenamefont {Moore}, \citenamefont {Johnston},
  \citenamefont {Cui}, \citenamefont {Li}, \citenamefont {Yi}, \citenamefont
  {Liu}, \citenamefont {Hashimoto}, \citenamefont {Zhang}, \citenamefont {Lu},
  \citenamefont {Devereaux}, \citenamefont {Lee},\ and\ \citenamefont
  {Shen}}]{Lee2014}%
  \BibitemOpen
  \bibfield  {author} {\bibinfo {author} {\bibfnamefont {J.~J.}\ \bibnamefont
  {Lee}}, \bibinfo {author} {\bibfnamefont {F.~T.}\ \bibnamefont {Schmitt}},
  \bibinfo {author} {\bibfnamefont {R.~G.}\ \bibnamefont {Moore}}, \bibinfo
  {author} {\bibfnamefont {S.}~\bibnamefont {Johnston}}, \bibinfo {author}
  {\bibfnamefont {Y.-T.}\ \bibnamefont {Cui}}, \bibinfo {author} {\bibfnamefont
  {W.}~\bibnamefont {Li}}, \bibinfo {author} {\bibfnamefont {M.}~\bibnamefont
  {Yi}}, \bibinfo {author} {\bibfnamefont {Z.~K.}\ \bibnamefont {Liu}},
  \bibinfo {author} {\bibfnamefont {M.}~\bibnamefont {Hashimoto}}, \bibinfo
  {author} {\bibfnamefont {Y.}~\bibnamefont {Zhang}}, \bibinfo {author}
  {\bibfnamefont {D.~H.}\ \bibnamefont {Lu}}, \bibinfo {author} {\bibfnamefont
  {T.~P.}\ \bibnamefont {Devereaux}}, \bibinfo {author} {\bibfnamefont {D.-H.}\
  \bibnamefont {Lee}}, \ and\ \bibinfo {author} {\bibfnamefont {Z.-X.}\
  \bibnamefont {Shen}},\ }\bibfield  {title} {\enquote {\bibinfo {title}
  {Interfacial mode coupling as the origin of the enhancement of {T$_c$} in
  {FeSe} films on {SrTiO$_3$}},}\ }\href {\doibase 10.1038/nature13894}
  {\bibfield  {journal} {\bibinfo  {journal} {Nature}\ }\textbf {\bibinfo
  {volume} {515}},\ \bibinfo {pages} {245--248} (\bibinfo {year}
  {2014})}\BibitemShut {NoStop}%
\bibitem [{\citenamefont {Song}\ \emph {et~al.}(2019)\citenamefont {Song},
  \citenamefont {Yu}, \citenamefont {Lou}, \citenamefont {Xie}, \citenamefont
  {Xu}, \citenamefont {Wen}, \citenamefont {Yao}, \citenamefont {Zhang},
  \citenamefont {Zhu}, \citenamefont {Guo}, \citenamefont {Peng},\ and\
  \citenamefont {Feng}}]{Song2019}%
  \BibitemOpen
  \bibfield  {author} {\bibinfo {author} {\bibfnamefont {Q.}~\bibnamefont
  {Song}}, \bibinfo {author} {\bibfnamefont {T.~L.}\ \bibnamefont {Yu}},
  \bibinfo {author} {\bibfnamefont {X.}~\bibnamefont {Lou}}, \bibinfo {author}
  {\bibfnamefont {B.~P.}\ \bibnamefont {Xie}}, \bibinfo {author} {\bibfnamefont
  {H.~C.}\ \bibnamefont {Xu}}, \bibinfo {author} {\bibfnamefont {C.~H.~P.}\
  \bibnamefont {Wen}}, \bibinfo {author} {\bibfnamefont {Q.}~\bibnamefont
  {Yao}}, \bibinfo {author} {\bibfnamefont {S.~Y.}\ \bibnamefont {Zhang}},
  \bibinfo {author} {\bibfnamefont {X.~T.}\ \bibnamefont {Zhu}}, \bibinfo
  {author} {\bibfnamefont {J.~D.}\ \bibnamefont {Guo}}, \bibinfo {author}
  {\bibfnamefont {R.}~\bibnamefont {Peng}}, \ and\ \bibinfo {author}
  {\bibfnamefont {D.~L.}\ \bibnamefont {Feng}},\ }\bibfield  {title} {\enquote
  {\bibinfo {title} {Evidence of cooperative effect on the enhanced
  superconducting transition temperature at the {FeSe}/{SrTiO$_3$}
  interface},}\ }\href {\doibase 10.1038/s41467-019-08560-z} {\bibfield
  {journal} {\bibinfo  {journal} {Nature Communications}\ }\textbf {\bibinfo
  {volume} {10}},\ \bibinfo {pages} {758} (\bibinfo {year} {2019})}\BibitemShut
  {NoStop}%
\bibitem [{\citenamefont {Yang}\ \emph
  {et~al.}(2015{\natexlab{b}})\citenamefont {Yang}, \citenamefont {Sobota},
  \citenamefont {Leuenberger}, \citenamefont {Kemper}, \citenamefont {Lee},
  \citenamefont {Schmitt}, \citenamefont {Li}, \citenamefont {Moore},
  \citenamefont {Kirchmann},\ and\ \citenamefont {Shen}}]{Yang2015thick}%
  \BibitemOpen
  \bibfield  {author} {\bibinfo {author} {\bibfnamefont {S.}~\bibnamefont
  {Yang}}, \bibinfo {author} {\bibfnamefont {J.~A.}\ \bibnamefont {Sobota}},
  \bibinfo {author} {\bibfnamefont {D.}~\bibnamefont {Leuenberger}}, \bibinfo
  {author} {\bibfnamefont {A.~F.}\ \bibnamefont {Kemper}}, \bibinfo {author}
  {\bibfnamefont {J.~J.}\ \bibnamefont {Lee}}, \bibinfo {author} {\bibfnamefont
  {F.~T.}\ \bibnamefont {Schmitt}}, \bibinfo {author} {\bibfnamefont
  {W.}~\bibnamefont {Li}}, \bibinfo {author} {\bibfnamefont {R.~G.}\
  \bibnamefont {Moore}}, \bibinfo {author} {\bibfnamefont {P.~S.}\ \bibnamefont
  {Kirchmann}}, \ and\ \bibinfo {author} {\bibfnamefont {Z.-X.}\ \bibnamefont
  {Shen}},\ }\bibfield  {title} {\enquote {\bibinfo {title}
  {Thickness-dependent coherent phonon frequency in ultrathin {FeSe/SrTiO$_3$}
  films},}\ }\href {\doibase 10.1021/acs.nanolett.5b01274} {\bibfield
  {journal} {\bibinfo  {journal} {Nano Letters}\ }\textbf {\bibinfo {volume}
  {15}},\ \bibinfo {pages} {4150--4154} (\bibinfo {year}
  {2015}{\natexlab{b}})}\BibitemShut {NoStop}%
\bibitem [{\citenamefont {Suzuki}\ \emph {et~al.}(2019)\citenamefont {Suzuki},
  \citenamefont {Someya}, \citenamefont {Hashimoto}, \citenamefont {Michimae},
  \citenamefont {Watanabe}, \citenamefont {Fujisawa}, \citenamefont {Kanai},
  \citenamefont {Ishii}, \citenamefont {Itatani}, \citenamefont {Kasahara},
  \citenamefont {Matsuda}, \citenamefont {Shibauchi}, \citenamefont {Okazaki},\
  and\ \citenamefont {Shin}}]{Suzuki2019}%
  \BibitemOpen
  \bibfield  {author} {\bibinfo {author} {\bibfnamefont {T.}~\bibnamefont
  {Suzuki}}, \bibinfo {author} {\bibfnamefont {T.}~\bibnamefont {Someya}},
  \bibinfo {author} {\bibfnamefont {T.}~\bibnamefont {Hashimoto}}, \bibinfo
  {author} {\bibfnamefont {S.}~\bibnamefont {Michimae}}, \bibinfo {author}
  {\bibfnamefont {M.}~\bibnamefont {Watanabe}}, \bibinfo {author}
  {\bibfnamefont {M.}~\bibnamefont {Fujisawa}}, \bibinfo {author}
  {\bibfnamefont {T.}~\bibnamefont {Kanai}}, \bibinfo {author} {\bibfnamefont
  {N.}~\bibnamefont {Ishii}}, \bibinfo {author} {\bibfnamefont
  {J.}~\bibnamefont {Itatani}}, \bibinfo {author} {\bibfnamefont
  {S.}~\bibnamefont {Kasahara}}, \bibinfo {author} {\bibfnamefont
  {Y.}~\bibnamefont {Matsuda}}, \bibinfo {author} {\bibfnamefont
  {T.}~\bibnamefont {Shibauchi}}, \bibinfo {author} {\bibfnamefont
  {K.}~\bibnamefont {Okazaki}}, \ and\ \bibinfo {author} {\bibfnamefont
  {S.}~\bibnamefont {Shin}},\ }\bibfield  {title} {\enquote {\bibinfo {title}
  {Photoinduced possible superconducting state with long-lived disproportionate
  band filling in {FeSe}},}\ }\href {\doibase 10.1038/s42005-019-0219-4}
  {\bibfield  {journal} {\bibinfo  {journal} {Communications Physics}\ }\textbf
  {\bibinfo {volume} {2}},\ \bibinfo {pages} {1--7} (\bibinfo {year}
  {2019})}\BibitemShut {NoStop}%
\bibitem [{\citenamefont {Zhang}\ \emph {et~al.}(2016)\citenamefont {Zhang},
  \citenamefont {Yi}, \citenamefont {Liu}, \citenamefont {Li}, \citenamefont
  {Lee}, \citenamefont {Moore}, \citenamefont {Hashimoto}, \citenamefont
  {Nakajima}, \citenamefont {Eisaki}, \citenamefont {Mo}, \citenamefont
  {Hussain}, \citenamefont {Devereaux}, \citenamefont {Shen},\ and\
  \citenamefont {Lu}}]{Zhang2016}%
  \BibitemOpen
  \bibfield  {author} {\bibinfo {author} {\bibfnamefont {Y.}~\bibnamefont
  {Zhang}}, \bibinfo {author} {\bibfnamefont {M.}~\bibnamefont {Yi}}, \bibinfo
  {author} {\bibfnamefont {Z.-K.}\ \bibnamefont {Liu}}, \bibinfo {author}
  {\bibfnamefont {W.}~\bibnamefont {Li}}, \bibinfo {author} {\bibfnamefont
  {J.~J.}\ \bibnamefont {Lee}}, \bibinfo {author} {\bibfnamefont {R.~G.}\
  \bibnamefont {Moore}}, \bibinfo {author} {\bibfnamefont {M.}~\bibnamefont
  {Hashimoto}}, \bibinfo {author} {\bibfnamefont {M.}~\bibnamefont {Nakajima}},
  \bibinfo {author} {\bibfnamefont {H.}~\bibnamefont {Eisaki}}, \bibinfo
  {author} {\bibfnamefont {S.-K.}\ \bibnamefont {Mo}}, \bibinfo {author}
  {\bibfnamefont {Z.}~\bibnamefont {Hussain}}, \bibinfo {author} {\bibfnamefont
  {T.~P.}\ \bibnamefont {Devereaux}}, \bibinfo {author} {\bibfnamefont {Z.-X.}\
  \bibnamefont {Shen}}, \ and\ \bibinfo {author} {\bibfnamefont {D.~H.}\
  \bibnamefont {Lu}},\ }\bibfield  {title} {\enquote {\bibinfo {title}
  {Distinctive orbital anisotropy observed in the nematic state of a {FeSe}
  thin film},}\ }\href {\doibase 10.1103/PhysRevB.94.115153} {\bibfield
  {journal} {\bibinfo  {journal} {Physical Review B}\ }\textbf {\bibinfo
  {volume} {94}},\ \bibinfo {pages} {115153} (\bibinfo {year}
  {2016})}\BibitemShut {NoStop}%
\bibitem [{\citenamefont {Yi}\ \emph {et~al.}(2019)\citenamefont {Yi},
  \citenamefont {Pfau}, \citenamefont {Zhang}, \citenamefont {He},
  \citenamefont {Wu}, \citenamefont {Chen}, \citenamefont {Ye}, \citenamefont
  {Hashimoto}, \citenamefont {Yu}, \citenamefont {Si}, \citenamefont {Lee},
  \citenamefont {Dai}, \citenamefont {Shen}, \citenamefont {Lu},\ and\
  \citenamefont {Birgeneau}}]{Yi2019}%
  \BibitemOpen
  \bibfield  {author} {\bibinfo {author} {\bibfnamefont {M.}~\bibnamefont
  {Yi}}, \bibinfo {author} {\bibfnamefont {H.}~\bibnamefont {Pfau}}, \bibinfo
  {author} {\bibfnamefont {Y.}~\bibnamefont {Zhang}}, \bibinfo {author}
  {\bibfnamefont {Y.}~\bibnamefont {He}}, \bibinfo {author} {\bibfnamefont
  {H.}~\bibnamefont {Wu}}, \bibinfo {author} {\bibfnamefont {T.}~\bibnamefont
  {Chen}}, \bibinfo {author} {\bibfnamefont {Z.~R.}\ \bibnamefont {Ye}},
  \bibinfo {author} {\bibfnamefont {M.}~\bibnamefont {Hashimoto}}, \bibinfo
  {author} {\bibfnamefont {R.}~\bibnamefont {Yu}}, \bibinfo {author}
  {\bibfnamefont {Q.}~\bibnamefont {Si}}, \bibinfo {author} {\bibfnamefont
  {D.-H.}\ \bibnamefont {Lee}}, \bibinfo {author} {\bibfnamefont
  {P.}~\bibnamefont {Dai}}, \bibinfo {author} {\bibfnamefont {Z.-X.}\
  \bibnamefont {Shen}}, \bibinfo {author} {\bibfnamefont {D.~H.}\ \bibnamefont
  {Lu}}, \ and\ \bibinfo {author} {\bibfnamefont {R.~J.}\ \bibnamefont
  {Birgeneau}},\ }\bibfield  {title} {\enquote {\bibinfo {title} {Nematic
  energy scale and the missing electron pocket in {FeSe}},}\ }\href {\doibase
  10.1103/PhysRevX.9.041049} {\bibfield  {journal} {\bibinfo  {journal}
  {Physical Review X}\ }\textbf {\bibinfo {volume} {9}},\ \bibinfo {pages}
  {041049} (\bibinfo {year} {2019})}\BibitemShut {NoStop}%
\bibitem [{\citenamefont {Huang}\ \emph {et~al.}(2021)\citenamefont {Huang},
  \citenamefont {Yu}, \citenamefont {Xu}, \citenamefont {Zhu}, \citenamefont
  {Jiang}, \citenamefont {Wang}, \citenamefont {Wu}, \citenamefont {Chen},
  \citenamefont {Denlinger}, \citenamefont {Mo}, \citenamefont {Hashimoto},
  \citenamefont {Gu}, \citenamefont {Dai}, \citenamefont {Chu}, \citenamefont
  {Lu}, \citenamefont {Si}, \citenamefont {Birgeneau},\ and\ \citenamefont
  {Yi}}]{huang2021}%
  \BibitemOpen
  \bibfield  {author} {\bibinfo {author} {\bibfnamefont {J.}~\bibnamefont
  {Huang}}, \bibinfo {author} {\bibfnamefont {R.}~\bibnamefont {Yu}}, \bibinfo
  {author} {\bibfnamefont {Z.}~\bibnamefont {Xu}}, \bibinfo {author}
  {\bibfnamefont {J.-X.}\ \bibnamefont {Zhu}}, \bibinfo {author} {\bibfnamefont
  {Q.}~\bibnamefont {Jiang}}, \bibinfo {author} {\bibfnamefont
  {M.}~\bibnamefont {Wang}}, \bibinfo {author} {\bibfnamefont {H.}~\bibnamefont
  {Wu}}, \bibinfo {author} {\bibfnamefont {T.}~\bibnamefont {Chen}}, \bibinfo
  {author} {\bibfnamefont {J.~D.}\ \bibnamefont {Denlinger}}, \bibinfo {author}
  {\bibfnamefont {S.-K.}\ \bibnamefont {Mo}}, \bibinfo {author} {\bibfnamefont
  {M.}~\bibnamefont {Hashimoto}}, \bibinfo {author} {\bibfnamefont
  {G.}~\bibnamefont {Gu}}, \bibinfo {author} {\bibfnamefont {P.}~\bibnamefont
  {Dai}}, \bibinfo {author} {\bibfnamefont {J.-H.}\ \bibnamefont {Chu}},
  \bibinfo {author} {\bibfnamefont {D.}~\bibnamefont {Lu}}, \bibinfo {author}
  {\bibfnamefont {Q.}~\bibnamefont {Si}}, \bibinfo {author} {\bibfnamefont
  {R.~J.}\ \bibnamefont {Birgeneau}}, \ and\ \bibinfo {author} {\bibfnamefont
  {M.}~\bibnamefont {Yi}},\ }\href@noop {} {\enquote {\bibinfo {title}
  {Non-thermal emergence of an orbital-selective {Mott} phase in
  {Fe}{Te$_{1-x}$}{Se$_x$}},}\ } (\bibinfo {year} {2021}),\ \Eprint
  {http://arxiv.org/abs/2010.13913} {arXiv:2010.13913} \BibitemShut {NoStop}%
\bibitem [{\citenamefont {Grévin}\ \emph {et~al.}(2011)\citenamefont
  {Grévin}, \citenamefont {Fakir}, \citenamefont {Hayton}, \citenamefont
  {Brun}, \citenamefont {Demadrille},\ and\ \citenamefont
  {Faure-Vincent}}]{Grevin2011}%
  \BibitemOpen
  \bibfield  {author} {\bibinfo {author} {\bibfnamefont {B.}~\bibnamefont
  {Grévin}}, \bibinfo {author} {\bibfnamefont {M.}~\bibnamefont {Fakir}},
  \bibinfo {author} {\bibfnamefont {J.}~\bibnamefont {Hayton}}, \bibinfo
  {author} {\bibfnamefont {M.}~\bibnamefont {Brun}}, \bibinfo {author}
  {\bibfnamefont {R.}~\bibnamefont {Demadrille}}, \ and\ \bibinfo {author}
  {\bibfnamefont {J.}~\bibnamefont {Faure-Vincent}},\ }\bibfield  {title}
  {\enquote {\bibinfo {title} {Qplus {AFM} driven nanostencil},}\ }\href
  {\doibase 10.1063/1.3600898} {\bibfield  {journal} {\bibinfo  {journal}
  {Review of Scientific Instruments}\ }\textbf {\bibinfo {volume} {82}},\
  \bibinfo {pages} {063706} (\bibinfo {year} {2011})}\BibitemShut {NoStop}%
\bibitem [{\citenamefont {Hu}\ \emph {et~al.}(2020{\natexlab{a}})\citenamefont
  {Hu}, \citenamefont {Xu}, \citenamefont {Shi}, \citenamefont {Luo},
  \citenamefont {Peng}, \citenamefont {Wang}, \citenamefont {Ying},
  \citenamefont {Wu}, \citenamefont {Liu}, \citenamefont {Zhang}, \citenamefont
  {Chen}, \citenamefont {Xu}, \citenamefont {Chen},\ and\ \citenamefont
  {He}}]{Hu2020}%
  \BibitemOpen
  \bibfield  {author} {\bibinfo {author} {\bibfnamefont {Y.}~\bibnamefont
  {Hu}}, \bibinfo {author} {\bibfnamefont {L.}~\bibnamefont {Xu}}, \bibinfo
  {author} {\bibfnamefont {M.}~\bibnamefont {Shi}}, \bibinfo {author}
  {\bibfnamefont {A.}~\bibnamefont {Luo}}, \bibinfo {author} {\bibfnamefont
  {S.}~\bibnamefont {Peng}}, \bibinfo {author} {\bibfnamefont {Z.~Y.}\
  \bibnamefont {Wang}}, \bibinfo {author} {\bibfnamefont {J.~J.}\ \bibnamefont
  {Ying}}, \bibinfo {author} {\bibfnamefont {T.}~\bibnamefont {Wu}}, \bibinfo
  {author} {\bibfnamefont {Z.~K.}\ \bibnamefont {Liu}}, \bibinfo {author}
  {\bibfnamefont {C.~F.}\ \bibnamefont {Zhang}}, \bibinfo {author}
  {\bibfnamefont {Y.~L.}\ \bibnamefont {Chen}}, \bibinfo {author}
  {\bibfnamefont {G.}~\bibnamefont {Xu}}, \bibinfo {author} {\bibfnamefont
  {X.-H.}\ \bibnamefont {Chen}}, \ and\ \bibinfo {author} {\bibfnamefont
  {J.-F.}\ \bibnamefont {He}},\ }\bibfield  {title} {\enquote {\bibinfo {title}
  {Universal gapless {Dirac} cone and tunable topological states in
  ${(\mathrm{MnB}{\mathrm{i}}_{2}\mathrm{T}{\mathrm{e}}_{4})}_{m}{(\mathrm{B}{\mathrm{i}}_{2}\mathrm{T}{\mathrm{e}}_{3})}_{n}$
  heterostructures},}\ }\href {\doibase 10.1103/PhysRevB.101.161113} {\bibfield
   {journal} {\bibinfo  {journal} {Physical Review B}\ }\textbf {\bibinfo
  {volume} {101}},\ \bibinfo {pages} {161113} (\bibinfo {year}
  {2020}{\natexlab{a}})}\BibitemShut {NoStop}%
\bibitem [{\citenamefont {Vidal}\ \emph {et~al.}(2021)\citenamefont {Vidal},
  \citenamefont {Bentmann}, \citenamefont {Facio}, \citenamefont {Heider},
  \citenamefont {Kagerer}, \citenamefont {Fornari}, \citenamefont {Peixoto},
  \citenamefont {Figgemeier}, \citenamefont {Jung}, \citenamefont {Cacho},
  \citenamefont {B\"uchner}, \citenamefont {van~den Brink}, \citenamefont
  {Schneider}, \citenamefont {Plucinski}, \citenamefont {Schwier},
  \citenamefont {Shimada}, \citenamefont {Richter}, \citenamefont {Isaeva},\
  and\ \citenamefont {Reinert}}]{Vidal2021}%
  \BibitemOpen
  \bibfield  {author} {\bibinfo {author} {\bibfnamefont {R.~C.}\ \bibnamefont
  {Vidal}}, \bibinfo {author} {\bibfnamefont {H.}~\bibnamefont {Bentmann}},
  \bibinfo {author} {\bibfnamefont {J.~I.}\ \bibnamefont {Facio}}, \bibinfo
  {author} {\bibfnamefont {T.}~\bibnamefont {Heider}}, \bibinfo {author}
  {\bibfnamefont {P.}~\bibnamefont {Kagerer}}, \bibinfo {author} {\bibfnamefont
  {C.~I.}\ \bibnamefont {Fornari}}, \bibinfo {author} {\bibfnamefont
  {T.~R.~F.}\ \bibnamefont {Peixoto}}, \bibinfo {author} {\bibfnamefont
  {T.}~\bibnamefont {Figgemeier}}, \bibinfo {author} {\bibfnamefont
  {S.}~\bibnamefont {Jung}}, \bibinfo {author} {\bibfnamefont {C.}~\bibnamefont
  {Cacho}}, \bibinfo {author} {\bibfnamefont {B.}~\bibnamefont {B\"uchner}},
  \bibinfo {author} {\bibfnamefont {J.}~\bibnamefont {van~den Brink}}, \bibinfo
  {author} {\bibfnamefont {C.~M.}\ \bibnamefont {Schneider}}, \bibinfo {author}
  {\bibfnamefont {L.}~\bibnamefont {Plucinski}}, \bibinfo {author}
  {\bibfnamefont {E.~F.}\ \bibnamefont {Schwier}}, \bibinfo {author}
  {\bibfnamefont {K.}~\bibnamefont {Shimada}}, \bibinfo {author} {\bibfnamefont
  {M.}~\bibnamefont {Richter}}, \bibinfo {author} {\bibfnamefont
  {A.}~\bibnamefont {Isaeva}}, \ and\ \bibinfo {author} {\bibfnamefont
  {F.}~\bibnamefont {Reinert}},\ }\bibfield  {title} {\enquote {\bibinfo
  {title} {Orbital complexity in intrinsic magnetic topological insulators
  {MnBi$_4$Te$_7$} and {MnBi$_6$Te$_{10}$}},}\ }\href {\doibase
  10.1103/PhysRevLett.126.176403} {\bibfield  {journal} {\bibinfo  {journal}
  {Physical Review Letters}\ }\textbf {\bibinfo {volume} {126}},\ \bibinfo
  {pages} {176403} (\bibinfo {year} {2021})}\BibitemShut {NoStop}%
\bibitem [{\citenamefont {Wu}\ \emph {et~al.}(2020)\citenamefont {Wu},
  \citenamefont {Li}, \citenamefont {Ma}, \citenamefont {Zhang}, \citenamefont
  {Liu}, \citenamefont {Zhou}, \citenamefont {Shao}, \citenamefont {Wang},
  \citenamefont {Hao}, \citenamefont {Feng}, \citenamefont {Schwier},
  \citenamefont {Kumar}, \citenamefont {Sun}, \citenamefont {Liu},
  \citenamefont {Shimada}, \citenamefont {Miyamoto}, \citenamefont {Okuda},
  \citenamefont {Wang}, \citenamefont {Xie}, \citenamefont {Chen},
  \citenamefont {Liu}, \citenamefont {Liu},\ and\ \citenamefont
  {Zhao}}]{Wu2020}%
  \BibitemOpen
  \bibfield  {author} {\bibinfo {author} {\bibfnamefont {X.}~\bibnamefont
  {Wu}}, \bibinfo {author} {\bibfnamefont {J.}~\bibnamefont {Li}}, \bibinfo
  {author} {\bibfnamefont {X.-M.}\ \bibnamefont {Ma}}, \bibinfo {author}
  {\bibfnamefont {Y.}~\bibnamefont {Zhang}}, \bibinfo {author} {\bibfnamefont
  {Y.}~\bibnamefont {Liu}}, \bibinfo {author} {\bibfnamefont {C.-S.}\
  \bibnamefont {Zhou}}, \bibinfo {author} {\bibfnamefont {J.}~\bibnamefont
  {Shao}}, \bibinfo {author} {\bibfnamefont {Q.}~\bibnamefont {Wang}}, \bibinfo
  {author} {\bibfnamefont {Y.-J.}\ \bibnamefont {Hao}}, \bibinfo {author}
  {\bibfnamefont {Y.}~\bibnamefont {Feng}}, \bibinfo {author} {\bibfnamefont
  {E.~F.}\ \bibnamefont {Schwier}}, \bibinfo {author} {\bibfnamefont
  {S.}~\bibnamefont {Kumar}}, \bibinfo {author} {\bibfnamefont
  {H.}~\bibnamefont {Sun}}, \bibinfo {author} {\bibfnamefont {P.}~\bibnamefont
  {Liu}}, \bibinfo {author} {\bibfnamefont {K.}~\bibnamefont {Shimada}},
  \bibinfo {author} {\bibfnamefont {K.}~\bibnamefont {Miyamoto}}, \bibinfo
  {author} {\bibfnamefont {T.}~\bibnamefont {Okuda}}, \bibinfo {author}
  {\bibfnamefont {K.}~\bibnamefont {Wang}}, \bibinfo {author} {\bibfnamefont
  {M.}~\bibnamefont {Xie}}, \bibinfo {author} {\bibfnamefont {C.}~\bibnamefont
  {Chen}}, \bibinfo {author} {\bibfnamefont {Q.}~\bibnamefont {Liu}}, \bibinfo
  {author} {\bibfnamefont {C.}~\bibnamefont {Liu}}, \ and\ \bibinfo {author}
  {\bibfnamefont {Y.}~\bibnamefont {Zhao}},\ }\bibfield  {title} {\enquote
  {\bibinfo {title} {Distinct topological surface states on the two
  terminations of {MnBi$_4$Te$_7$}},}\ }\href {\doibase
  10.1103/PhysRevX.10.031013} {\bibfield  {journal} {\bibinfo  {journal}
  {Physical Review X}\ }\textbf {\bibinfo {volume} {10}},\ \bibinfo {pages}
  {031013} (\bibinfo {year} {2020})}\BibitemShut {NoStop}%
\bibitem [{\citenamefont {Hu}\ \emph {et~al.}(2020{\natexlab{b}})\citenamefont
  {Hu}, \citenamefont {Gordon}, \citenamefont {Liu}, \citenamefont {Liu},
  \citenamefont {Zhou}, \citenamefont {Hao}, \citenamefont {Narayan},
  \citenamefont {Emmanouilidou}, \citenamefont {Sun}, \citenamefont {Liu},
  \citenamefont {Brawer}, \citenamefont {Ramirez}, \citenamefont {Ding},
  \citenamefont {Cao}, \citenamefont {Liu}, \citenamefont {Dessau},\ and\
  \citenamefont {Ni}}]{Hu2020a}%
  \BibitemOpen
  \bibfield  {author} {\bibinfo {author} {\bibfnamefont {C.}~\bibnamefont
  {Hu}}, \bibinfo {author} {\bibfnamefont {K.~N.}\ \bibnamefont {Gordon}},
  \bibinfo {author} {\bibfnamefont {P.}~\bibnamefont {Liu}}, \bibinfo {author}
  {\bibfnamefont {J.}~\bibnamefont {Liu}}, \bibinfo {author} {\bibfnamefont
  {X.}~\bibnamefont {Zhou}}, \bibinfo {author} {\bibfnamefont {P.}~\bibnamefont
  {Hao}}, \bibinfo {author} {\bibfnamefont {D.}~\bibnamefont {Narayan}},
  \bibinfo {author} {\bibfnamefont {E.}~\bibnamefont {Emmanouilidou}}, \bibinfo
  {author} {\bibfnamefont {H.}~\bibnamefont {Sun}}, \bibinfo {author}
  {\bibfnamefont {Y.}~\bibnamefont {Liu}}, \bibinfo {author} {\bibfnamefont
  {H.}~\bibnamefont {Brawer}}, \bibinfo {author} {\bibfnamefont {A.~P.}\
  \bibnamefont {Ramirez}}, \bibinfo {author} {\bibfnamefont {L.}~\bibnamefont
  {Ding}}, \bibinfo {author} {\bibfnamefont {H.}~\bibnamefont {Cao}}, \bibinfo
  {author} {\bibfnamefont {Q.}~\bibnamefont {Liu}}, \bibinfo {author}
  {\bibfnamefont {D.}~\bibnamefont {Dessau}}, \ and\ \bibinfo {author}
  {\bibfnamefont {N.}~\bibnamefont {Ni}},\ }\bibfield  {title} {\enquote
  {\bibinfo {title} {A van der {Waals} antiferromagnetic topological insulator
  with weak interlayer magnetic coupling},}\ }\href {\doibase
  10.1038/s41467-019-13814-x} {\bibfield  {journal} {\bibinfo  {journal}
  {Nature Communications}\ }\textbf {\bibinfo {volume} {11}},\ \bibinfo {pages}
  {97} (\bibinfo {year} {2020}{\natexlab{b}})}\BibitemShut {NoStop}%
\bibitem [{\citenamefont {Ma}\ \emph {et~al.}(2020)\citenamefont {Ma},
  \citenamefont {Chen}, \citenamefont {Schwier}, \citenamefont {Zhang},
  \citenamefont {Hao}, \citenamefont {Kumar}, \citenamefont {Lu}, \citenamefont
  {Shao}, \citenamefont {Jin}, \citenamefont {Zeng}, \citenamefont {Liu},
  \citenamefont {Hao}, \citenamefont {Zhang}, \citenamefont {Mansuer},
  \citenamefont {Song}, \citenamefont {Wang}, \citenamefont {Zhao},
  \citenamefont {Liu}, \citenamefont {Deng}, \citenamefont {Mei}, \citenamefont
  {Shimada}, \citenamefont {Zhao}, \citenamefont {Zhou}, \citenamefont {Shen},
  \citenamefont {Huang}, \citenamefont {Liu}, \citenamefont {Xu},\ and\
  \citenamefont {Chen}}]{Ma2020}%
  \BibitemOpen
  \bibfield  {author} {\bibinfo {author} {\bibfnamefont {X.-M.}\ \bibnamefont
  {Ma}}, \bibinfo {author} {\bibfnamefont {Z.}~\bibnamefont {Chen}}, \bibinfo
  {author} {\bibfnamefont {E.~F.}\ \bibnamefont {Schwier}}, \bibinfo {author}
  {\bibfnamefont {Y.}~\bibnamefont {Zhang}}, \bibinfo {author} {\bibfnamefont
  {Y.-J.}\ \bibnamefont {Hao}}, \bibinfo {author} {\bibfnamefont
  {S.}~\bibnamefont {Kumar}}, \bibinfo {author} {\bibfnamefont
  {R.}~\bibnamefont {Lu}}, \bibinfo {author} {\bibfnamefont {J.}~\bibnamefont
  {Shao}}, \bibinfo {author} {\bibfnamefont {Y.}~\bibnamefont {Jin}}, \bibinfo
  {author} {\bibfnamefont {M.}~\bibnamefont {Zeng}}, \bibinfo {author}
  {\bibfnamefont {X.-R.}\ \bibnamefont {Liu}}, \bibinfo {author} {\bibfnamefont
  {Z.}~\bibnamefont {Hao}}, \bibinfo {author} {\bibfnamefont {K.}~\bibnamefont
  {Zhang}}, \bibinfo {author} {\bibfnamefont {W.}~\bibnamefont {Mansuer}},
  \bibinfo {author} {\bibfnamefont {C.}~\bibnamefont {Song}}, \bibinfo {author}
  {\bibfnamefont {Y.}~\bibnamefont {Wang}}, \bibinfo {author} {\bibfnamefont
  {B.}~\bibnamefont {Zhao}}, \bibinfo {author} {\bibfnamefont {C.}~\bibnamefont
  {Liu}}, \bibinfo {author} {\bibfnamefont {K.}~\bibnamefont {Deng}}, \bibinfo
  {author} {\bibfnamefont {J.}~\bibnamefont {Mei}}, \bibinfo {author}
  {\bibfnamefont {K.}~\bibnamefont {Shimada}}, \bibinfo {author} {\bibfnamefont
  {Y.}~\bibnamefont {Zhao}}, \bibinfo {author} {\bibfnamefont {X.}~\bibnamefont
  {Zhou}}, \bibinfo {author} {\bibfnamefont {B.}~\bibnamefont {Shen}}, \bibinfo
  {author} {\bibfnamefont {W.}~\bibnamefont {Huang}}, \bibinfo {author}
  {\bibfnamefont {C.}~\bibnamefont {Liu}}, \bibinfo {author} {\bibfnamefont
  {H.}~\bibnamefont {Xu}}, \ and\ \bibinfo {author} {\bibfnamefont
  {C.}~\bibnamefont {Chen}},\ }\bibfield  {title} {\enquote {\bibinfo {title}
  {Hybridization-induced gapped and gapless states on the surface of magnetic
  topological insulators},}\ }\href {\doibase 10.1103/PhysRevB.102.245136}
  {\bibfield  {journal} {\bibinfo  {journal} {Physical Review B}\ }\textbf
  {\bibinfo {volume} {102}},\ \bibinfo {pages} {245136} (\bibinfo {year}
  {2020})}\BibitemShut {NoStop}%
\bibitem [{\citenamefont {Yan}\ \emph {et~al.}(2021{\natexlab{a}})\citenamefont
  {Yan}, \citenamefont {Fernandez-Mulligan}, \citenamefont {Mei}, \citenamefont
  {Lee}, \citenamefont {Protic}, \citenamefont {Fukumori}, \citenamefont {Yan},
  \citenamefont {Liu}, \citenamefont {Mao},\ and\ \citenamefont
  {Yang}}]{Yan2021}%
  \BibitemOpen
  \bibfield  {author} {\bibinfo {author} {\bibfnamefont {C.}~\bibnamefont
  {Yan}}, \bibinfo {author} {\bibfnamefont {S.}~\bibnamefont
  {Fernandez-Mulligan}}, \bibinfo {author} {\bibfnamefont {R.}~\bibnamefont
  {Mei}}, \bibinfo {author} {\bibfnamefont {S.~H.}\ \bibnamefont {Lee}},
  \bibinfo {author} {\bibfnamefont {N.}~\bibnamefont {Protic}}, \bibinfo
  {author} {\bibfnamefont {R.}~\bibnamefont {Fukumori}}, \bibinfo {author}
  {\bibfnamefont {B.}~\bibnamefont {Yan}}, \bibinfo {author} {\bibfnamefont
  {C.}~\bibnamefont {Liu}}, \bibinfo {author} {\bibfnamefont {Z.}~\bibnamefont
  {Mao}}, \ and\ \bibinfo {author} {\bibfnamefont {S.}~\bibnamefont {Yang}},\
  }\bibfield  {title} {\enquote {\bibinfo {title} {Origins of electronic bands
  in the antiferromagnetic topological insulator {MnBi$_2$Te$_4$}},}\ }\href
  {\doibase 10.1103/PhysRevB.104.L041102} {\bibfield  {journal} {\bibinfo
  {journal} {Physical Review B}\ }\textbf {\bibinfo {volume} {104}},\ \bibinfo
  {pages} {L041102} (\bibinfo {year} {2021}{\natexlab{a}})}\BibitemShut
  {NoStop}%
\bibitem [{\citenamefont {Yan}\ \emph {et~al.}(2021{\natexlab{b}})\citenamefont
  {Yan}, \citenamefont {Zhu}, \citenamefont {Fernandez-Mulligan}, \citenamefont
  {Green}, \citenamefont {Mei}, \citenamefont {Yan}, \citenamefont {Liu},
  \citenamefont {Mao},\ and\ \citenamefont {Yang}}]{Yan2021a}%
  \BibitemOpen
  \bibfield  {author} {\bibinfo {author} {\bibfnamefont {C.}~\bibnamefont
  {Yan}}, \bibinfo {author} {\bibfnamefont {Y.}~\bibnamefont {Zhu}}, \bibinfo
  {author} {\bibfnamefont {S.}~\bibnamefont {Fernandez-Mulligan}}, \bibinfo
  {author} {\bibfnamefont {E.}~\bibnamefont {Green}}, \bibinfo {author}
  {\bibfnamefont {R.}~\bibnamefont {Mei}}, \bibinfo {author} {\bibfnamefont
  {B.}~\bibnamefont {Yan}}, \bibinfo {author} {\bibfnamefont {C.}~\bibnamefont
  {Liu}}, \bibinfo {author} {\bibfnamefont {Z.}~\bibnamefont {Mao}}, \ and\
  \bibinfo {author} {\bibfnamefont {S.}~\bibnamefont {Yang}},\ }\href@noop {}
  {\enquote {\bibinfo {title} {Delicate ferromagnetism in
  {MnBi$_6$Te$_{10}$}},}\ } (\bibinfo {year} {2021}{\natexlab{b}}),\ \Eprint
  {http://arxiv.org/abs/2107.08137} {arXiv:2107.08137} \BibitemShut {NoStop}%
\end{thebibliography}%

\end{document}